\documentclass[prd,twocolumn,fleqn,showpacs,showkeys,amsmath]{revtex4}

\usepackage{amsfonts}
\usepackage{graphicx}
\usepackage{psfrag}
\usepackage{amssymb}
\usepackage{color}
\usepackage{amsmath}

\newcommand{\rmi}{\mathrm{i}}
\newcommand{\rmd}{\mathrm{d}}

\newcommand{\bg}{\boldsymbol}
\renewcommand{\qquad}{\hspace*{25pt}}

\begin{document}

\setcounter{page}{1}%

\title{Macroscopic States in Bose--Einstein Condensate Dark Matter Model\\ with Axionlike Interaction}

\author{A.V.~Nazarenko\footnote{e-mail: nazarenko@bitp.kyiv.ua}}
\affiliation{Bogolyubov Institute for Theoretical Physics of NAS of Ukraine, \\ 
14b, Metrolohichna Str., Kyiv 03143, Ukraine}%


\begin{abstract}
The phase diagrams of ultralight dark matter (DM), modeled as a self-gravitating
Bose--Einstein condensate with axionlike interaction, are studied. We classify stable,
metastable, and unstable DM states over a wide range of condensate wave function
amplitudes. It is shown that the axionlike interaction causes instability and
an imaginary speed of sound at low amplitudes, whereas, in a specific
high-amplitude band, DM attains a stable state capable of forming a dense solitonic
core and suppressing quantum fluctuations in the surrounding galactic DM halo.
These findings are corroborated by evaluating thermodynamic functions for DM in
the dwarf galaxy NGC~2366 and its hypothetical analogs with different
core-to-halo mass ratios. Distinct DM phase compositions respond differently to
fluctuation-induced partial pressure, resulting in a first-order phase transition
in a certain range of an interaction parameter. While the DM properties in
NGC~2366 lie within the supercritical regime, the phase transition nonetheless
provides a thermodynamic marker separating stable from unstable DM configurations.
Once a dense core forms--reaching a threshold of about 12\% of 
the total mass--the enhanced gravitation stabilizes the DM halo against
fluctuations, while the internal pressure ensures core stability.
In particular, we find that NGC~2366's dense DM comprises roughly
19\% of the DM mass while occupying only 4.7\% of its total volume.
\end{abstract}



\maketitle

\section{Introduction}\label{sec1}

Several decades ago, axions were proposed as hypothetical pseudo-Goldstone
bosons~\cite{PQ77}, offering a solution to the charge-parity (CP) problem
in quantum chromodynamics (QCD). This innovative idea required an explanation
of both the emergence of axions during the Universe's evolution and their
broader cosmological implications. Axions were soon regarded as promising
candidates for dark matter (DM)~\cite{PWW}. The physical picture was
theoretically supported by mechanisms for their 
production~\cite{AS83,DF83,Dav86} and their transition to the nonrelativistic
condensate state~\cite{SY09,SY12,Dav13,Dav15,GHP15}. The latter allowed to
describe low-energy axions with large occupation numbers either as
a classical field in effective field theory or as a macroscopic wave function
in the Gross--Pitaevskii approach.

Often considered within the context of DM structures, modeling of axion stars
and clumps has revealed that their masses and sizes are relatively small from
a cosmological perspective (see \cite{BZ19}) and are constrained by the gravitational
stability conditions inherent to cold bosonic matter~\cite{BGZ84,UTB02}.
As argued in \cite{Bor16,KM17,FMT08,Ch2018}, the masses $10^{-4\pm1}~\text{eV}c^{-2}$
of axions preclude forming giant condensates comparable in size
to galactic DM halos.

In contrast, early attempts to model DM halos and rotation curves of galaxies
yielded extremely small values of the boson masses~\cite{Bal83,Sin} that are
tens of orders of magnitude smaller than those predicted in QCD.
Therefore, reproducing the observed data requires invoking dark ultralight
axions or axionlike particles~\cite{Mar16}.

Although the presence of self-interactions between ultralight DM particles
ensures a better description of astrophysical phenomena,
the fuzzy DM models often omit it (see \cite{Hu,Schive1,Chen17,Zim25}),
limiting themselves to taking into account only gravitational interactions,
as was originally assumed~\cite{Zwicky}.

It seems natural for us to investigate self-gravitating DM in real visible
objects using dark bosons with masses $10^{-22\pm1}~\text{eV}c^{-2}$,
incorporating the simplest instanton self-interaction borrowed from
QCD~\cite{BZ19}. Specifically, our goal is to apply the equilibrium
statistical approach to the DM in dwarf galaxies, effectively replacing,
in an ergodic sense, the analysis of the time evolution of a single
spatial configuration of DM by studying a statistical ensemble of
static configurations, which is governed by thermodynamic potential.
In this way we intend to classify and describe the phases of such a
matter.

Axionlike interactions can give rise to both stable and unstable states,
allowing their coexistence in nonuniform DM. It is already known that
the unstable state arises in the self-attraction mode~\cite{Nambo24} at small amplitudes
of the axion field, corresponding to rarefied matter. Although high amplitudes
seem unlikely for an axion field (but not necessarily for other
substances), some of their values should correspond to stable states~\cite{KT94}.
In any case, the overall state of the DM is determined by the spatial
distribution and relative abundance of these states, which collectively
define DM composition. Accordingly, the complete configurations are also
classified as either stable or unstable, and it is assumed that these regimes
can change via a first-order quantum phase transition at zero temperature.

One may observe parallels with the ensemble of oscillons~\cite{Osc1} generated
by the instanton self-interaction, leading in the absence of gravity to the
three-dimensional sine-Gordon equation~\cite{Man04}. Moreover, established
results on {\it axitons}~\cite{BZ19,KT94,SH2018} -- localized,
non-topological solitonic configurations of an axion or ultralight scalar field
stabilized by significant self-interactions -- provide a valuable benchmark for
comparing different families of field states.

Neglecting the Universe's evolution in describing current visible structures,
we are thus dealing with the statistical properties of self-interacting DM (SIDM)
within the model of self-gravitating Bose--Einstein condensate (BEC). Multifaceted
phase diagrams should then reflect these properties, tracing transitions among
distinct states and configurations.

Numerous studies have rigorously examined the properties of BEC DM
with two- and three-particle interactions, using both exact numerical
methods and various analytical approximations, such as the Thomas--Fermi
approximation and the variational method
(see \cite{Ch2018,Peebles,Bohmer,Ch2011,Harko2011,Zhang2018,Naz2020,SV24}).
The thermodynamic functions in these models have been carefully studied,
and a new aspect of DM has emerged~\cite{GKN20,GN21,GN23}: the prediction
of a first-order phase transition between rarefied and dense states, which
is driven by quantum fluctuation pressure and governed by coupling
constants. 
Additionally, we would like to note the impact of the formation
of composites from the initial DM particles~\cite{GN23,GN22,GKKN},
the quantum entanglement of parts of the DM halo~\cite{GN21,Lee18}, and
the emergence of superfluidity~\cite{Berezhiani,Ber25}.

Building on those studies, we transition from polynomial-type
self-interaction (over a field or wave function) to a nonlinear periodic
one.

To constrain the parameters of our BEC DM model described by
the Gross--Pitaevskii--Poisson equations, we consider the DM
dominating the dwarf galaxy NGC~2366 as a reference. This nearby galaxy
provides an opportunity to study reionization processes in vigorous
low-metallicity starbursts, similar to those that drove cosmic
reionization~\cite{Hun01,Iz99}, and exhibits the DM halo profile
inconsistent with the cuspy distributions predicted by cold DM
models ($\Lambda$CDM)~\cite{Oh08}. Earlier efforts have already
addressed this core-cusp tension within the BEC DM model with pairwise
interaction~\cite{Harko2011}. In this regard, we intend to calibrate
our model by fitting its rotation curve~\cite{Oh08,Bl08},
and then generate a statistical ensemble of hypothetical DM distributions
following from the model equations under similar conditions.
By comparing different configurations using phase diagrams,
we aim to reveal the uniqueness of the DM structure in NGC~2366,
which forms the gravitational background for subgalactic processes.
This analysis is expected to provide new insights into the galaxy's DM
and to test the diagnostic capabilities of the statistical approach.

The paper is structured as follows. In Sec.~\ref{S2} we introduce the main
equations of the model and, through various simplifications, analyze
the properties essential for interpreting subsequent results.
In Sec.~\ref{S3} we explore nonperturbative solutions and their application
to modeling the DM distribution in the dwarf galaxy NGC~2366.
We then present our main results -- statistical characteristics and
phase diagrams based on the numerically obtained DM profiles.
The final Sec.~\ref{S4} contains a summary of the findings
and concluding remarks.

\section{The model}\label{S2}
\subsection{The model equations}\label{S2A}

Consider a stationary Bose--Einstein condensate (BEC) of nonrelativistic
particles with mass~$m$, described by a macroscopic real wave
function $\psi(r)$ of a radial coordinate $r$ within a three-dimensional
ball $B(r\leq R)$. In the absence of hydrodynamic flows, the incorporation
of the axionlike self-interaction~$V_{\rm int}$ and the potential
of gravitational interaction~$V_{\rm gr}$ leads to the energy functional
$\Gamma=E-{\tilde\mu}{\cal N}$ with a constant chemical potential $\tilde\mu$:
\begin{eqnarray}
&&\Gamma=4\pi\int_0^R \widetilde{\cal F} r^2\,\rmd r,
\label{G1}\\
&&
\widetilde{\cal F}=\frac{\hbar^2}{2m}(\partial_r\psi)^2
+mV_{\rm gr}\psi^2+V_{\rm int}-\tilde\mu\psi^2,
\label{G11}\\
&&V_{\rm int}=\frac{U}{v}\left[1-\cos{\left(\sqrt{v}\psi\right)}\right]
-\frac{U}{2}\psi^2,
\label{SI1}\\
&&\Delta_r V_{\rm gr}=4\pi Gm\psi^2.
\label{Poi}
\end{eqnarray}

The spatial evolution of $\psi(r)$ is governed by the minimization
of the functional $\Gamma$ and the Poisson equation (\ref{Poi}).
This equation, along with its solution, is based on the radial
part of the Laplace operator $\Delta_r$ and its inverse $\Delta^{-1}_r$
with respect to the variable $r$:
\begin{eqnarray}
&&\hspace*{-5mm}
\Delta_rf(r)=\partial^2_rf(r)+\frac{2}{r}\,\partial_rf(r),
\label{Dlt}\\
&&\hspace*{-5mm}
\Delta^{-1}_rf(r)=-\frac{1}{r}\int_0^rf(s)s^2\rmd s-\int_r^{R}f(s)s\rmd s.
\label{InvDelta}
\end{eqnarray}

The axionlike interaction used here is chosen in analogy with \cite{PQ77,Wit80},
assuming the connection between the axion field $\varphi$ with the nonrelativistic
wave function $\psi$ via the relation $|\varphi|^2=(\hbar^2/m) |\psi|^2$~\cite{Ch2018}.
Other effective axion potentials also appear in \cite{VV80,Guzman,Cor16,Ch2017}.

Expanding (\ref{SI1}) in a series, we observe that the first term $\psi^4$
corresponds to a two-particle self-interaction, while the subsequent term $\psi^6$
indicates a three-particle self-interaction. The BEC DM models with only pairwise
repulsive interaction at $v=-|v|$ have been intensively studied in the Thomas--Fermi
approximation (e.g. see \cite{Ch2011,Harko2011,Zhang2018,Naz2020}).
Considering the three-particle interaction alongside quantum fluctuations
substantially enriches the physical picture, leading to a two-phase structure
and a first-order phase transition~\cite{GKN20,GN21}.

The axionlike interaction (\ref{SI1}) is characterized by two constants, $U$
and $v$, which have dimensions of energy and volume. In the particle
physics~\cite{VV80,Cor16}, they are related with the axion mass $m$ and
decay constant $f_{\text{a}}$ as $U=mc^2$, $v=\hbar^3c/(mf^2_{\text{a}})$.
Although the applicability of these relativistic relations in
nonrelativistic models is controversial, they allow us to make
comparisons.

In cosmological models, axions are described through various
parameterizations~\cite{Ch2018}. As highlighted in \cite{FMT08},
the choice of a smaller particle mass ensures the formation of certain
structures in the Universe, some of which we aim to describe.
It is expected that $m\sim10^{-22}~\text{eV} c^{-2}$, which is
usually attributed to the fuzzy DM~\cite{Hu,Peebles}.

Then, introducing the distance scale $r_0$, in nonrelativistic
quantum picture we fix $U$ as
\begin{equation}
U=\frac{\hbar^2}{mr_0^2}.
\end{equation}
Although the relativistic value $U=mc^2$ is recovered at the Compton wavelength
$r_0=\hbar/(mc)$, the chosen $U$ allows equalizing the scales of quantum
fluctuations and self-interaction, similar to the sine-Gordon model~\cite{Man04,Wil18}.
In any case, the implications of changing $U$ are discussed below.

This prescription immediately leads to the dimensionless measure of gravitational interaction:
\begin{equation}
A=\frac{8\pi Gm^3r_0^4}{\hbar^2v}\ll 1.
\end{equation}

With $v=\hbar^3c/(mf^2_{\text{a}})$ defined via the axion decay constant
$f_{\mathrm{a}}$, it is then expected that in models featuring ultralight
particles, $f_{\text{a}}$ may fall within the range 
$10^{18}\,\text{eV}<f_{\text{a}}<10^{21}\,\text{eV}$~\cite{Ch2018}.

Defining the dimensionless radial variable $\xi$ and wave function $\chi$,
\begin{equation}\label{mp1}
\xi=\frac{r}{r_0},\qquad
\chi(\xi)=\sqrt{v}\,\psi(r),
\end{equation}
we reduce the problem (\ref{G1})-(\ref{Poi}) to the following:
\begin{eqnarray}
&&\hspace*{-5mm}
\Gamma=\Gamma_0\int_0^{\xi_B} {\cal F} \xi^2\,\rmd\xi,\quad
\Gamma_0=\frac{2\pi r_0f^2_{\text{a}}}{\hbar c},
\label{feq3}\\
&&\hspace*{-5mm}
{\cal F}=(\partial_\xi\chi)^2
+A\Phi\chi^2+2(1-\cos{\chi})-\nu\chi^2,
\\
&&\hspace*{-5mm}
\Delta_\xi\Phi=\chi^2,\quad \Phi(0)=0.
\label{feq4}
\end{eqnarray}
It operates on a set $r_0$, $f_\text{a}$, $A$ and $\nu$,
such that
\begin{eqnarray}
&&m=\left[\frac{\hbar^5c}{8\pi G}\frac{A}{r_0^4f^2_\text{a}}\right]^{1/4},
\quad
\rho_0=\frac{m}{v},
\label{mass}\\
&&\nu=1+2u-A\Phi_0,\hspace*{8mm}
u=\frac{mr_0^2}{\hbar^2}\tilde\mu,
\end{eqnarray}
where $\rho_0$ is the mass density scale, and $\nu$ plays the role
of the effective chemical potential, which absorbs part of the axion
interaction and the gravitational potential at the origin $\xi=0$,
namely
\begin{equation}\label{f0}
\Phi_0=-\int_0^{\xi_B}\chi^2(\xi)\,\xi\,\rmd\xi.
\end{equation}
In our study, $\nu$ is regarded as a free parameter,
reflecting the arbitrariness of $\tilde\mu$.

Thus, the DM characteristics are
\begin{eqnarray}
m&=&0.17745\times10^{-22}~\text{eV}~c^{-2}~\left[\frac{A}{10^{-3}}\right]^{1/4}
\nonumber\\
&&\times\left[\frac{r_0}{1~\text{kpc}}\right]^{-1}
\left[\frac{f_\text{a}}{10^{19}~\text{eV}}\right]^{-1/2},\\
\rho_0&=&0.73055\times10^{-23}~\text{kg}~\text{m}^{-3}
~\left[\frac{A}{10^{-3}}\right]^{1/2}
\nonumber\\
&&\times\left[\frac{r_0}{1~\text{kpc}}\right]^{-2}
\left[\frac{f_\text{a}}{10^{19}~\text{eV}}\right],
\end{eqnarray}
so that the central mass density $\rho_c=\rho_0\chi^2(0)$.

The energy scale is defined as $\varepsilon_0=\hbar^2/(2mr_0^2)$. Combining
this with the mass scale, we get
\begin{eqnarray}
\varepsilon_0&=&1.1522\times10^{-30}~\text{eV}~\left[\frac{A}{10^{-3}}\right]^{-1/4}
\nonumber\\
&&\times\left[\frac{r_0}{1~\text{kpc}}\right]^{-1}
\left[\frac{f_\text{a}}{10^{19}~\text{eV}}\right]^{1/2}.
\end{eqnarray}

After the manipulations carried out, we come to the model equations in terms
of the wave-function $\chi(\xi)$ and gravitational potential $\Phi(\xi)$:
\begin{eqnarray}
&&\hspace*{-6mm}
\left(\Delta_{\xi}+\nu\right)\chi-A\Phi\chi-\sin{\chi}=0,
\label{feq1}\\
&&\hspace*{-6mm}
\Phi(\xi)=-\frac{1}{\xi}\int_0^\xi \chi^2(s)s^2\,\rmd s+\int_0^{\xi}\chi^2(s)s\,\rmd s,
\label{feq2}
\end{eqnarray}
where $\Delta_\xi\Phi=\chi^2$ is guaranteed, and $\Phi(0)=0$.

Note the scaling symmetry of Eqs.~(\ref{feq1})-(\ref{feq2}).
Let $\widetilde{\chi}(\xi)$ be a solution to the set of equations:
\begin{eqnarray}
&&\hspace*{-6mm}
\left(\Delta_{\xi}+\widetilde{\nu}\right)\widetilde{\chi}-\widetilde{A}\Phi\widetilde{\chi}-\widetilde{B}\sin{\widetilde{\chi}}=0,
\label{feq12}\\
&&\hspace*{-6mm}
\Phi(\xi)=-\frac{1}{\xi}\int_0^\xi \widetilde{\chi}^2(s)s^2\,\rmd s+\int_0^{\xi}\widetilde{\chi}^2(s)s\,\rmd s,
\label{feq22}
\end{eqnarray}

Suppose there exists a constant $\alpha$ such that $\widetilde{\nu}=\nu\alpha^2$,
$\widetilde{B}=\alpha^2$ and $\widetilde{A}=A\alpha^4$. Then, provided that 
the initial conditions agree $\widetilde{\chi}(0)=\chi(0)$, the solutions to
Eqs.~(\ref{feq1})-(\ref{feq2}) and (\ref{feq12})-(\ref{feq22}) are
related by $\widetilde{\chi}(\alpha\xi)=\chi(\xi)$. This means that replacing
any $\widetilde{B}>0$ by one changes the scale of $\xi$ and hence $r_0$,
which partially justifies using only the parameters $\nu$ and $A$ as independent.

In what follows we will focus on the analysis and solution of Eqs.~(\ref{feq1})-(\ref{feq2}),
where the first of them serves as the stationary Gross--Pitaevskii equation.

\subsection{\label{S2B}The particular cases}

{\bf Uniform states.}
Let us examine the simplest scenario -- a nonfluctuating
BEC DM without self-gravity. In this case, when $\partial_\xi\chi=0$ and
$A=0$, the grand canonical potential density (i.e. per unit volume) is
\begin{equation}\label{gom}
\omega=2(1-\cos{\chi})-\nu \chi^2,
\end{equation}
where $\nu$ is a chemical potential or external field.

The particle number density $\eta$ and volume per particle $v$ are
\begin{equation}
\eta\equiv-\frac{\partial\omega}{\partial\nu}=\chi^2,\quad
v=\frac{1}{\eta}.
\end{equation}

\begin{figure}[h]
\centering
\includegraphics[width=4.7cm,angle=0]{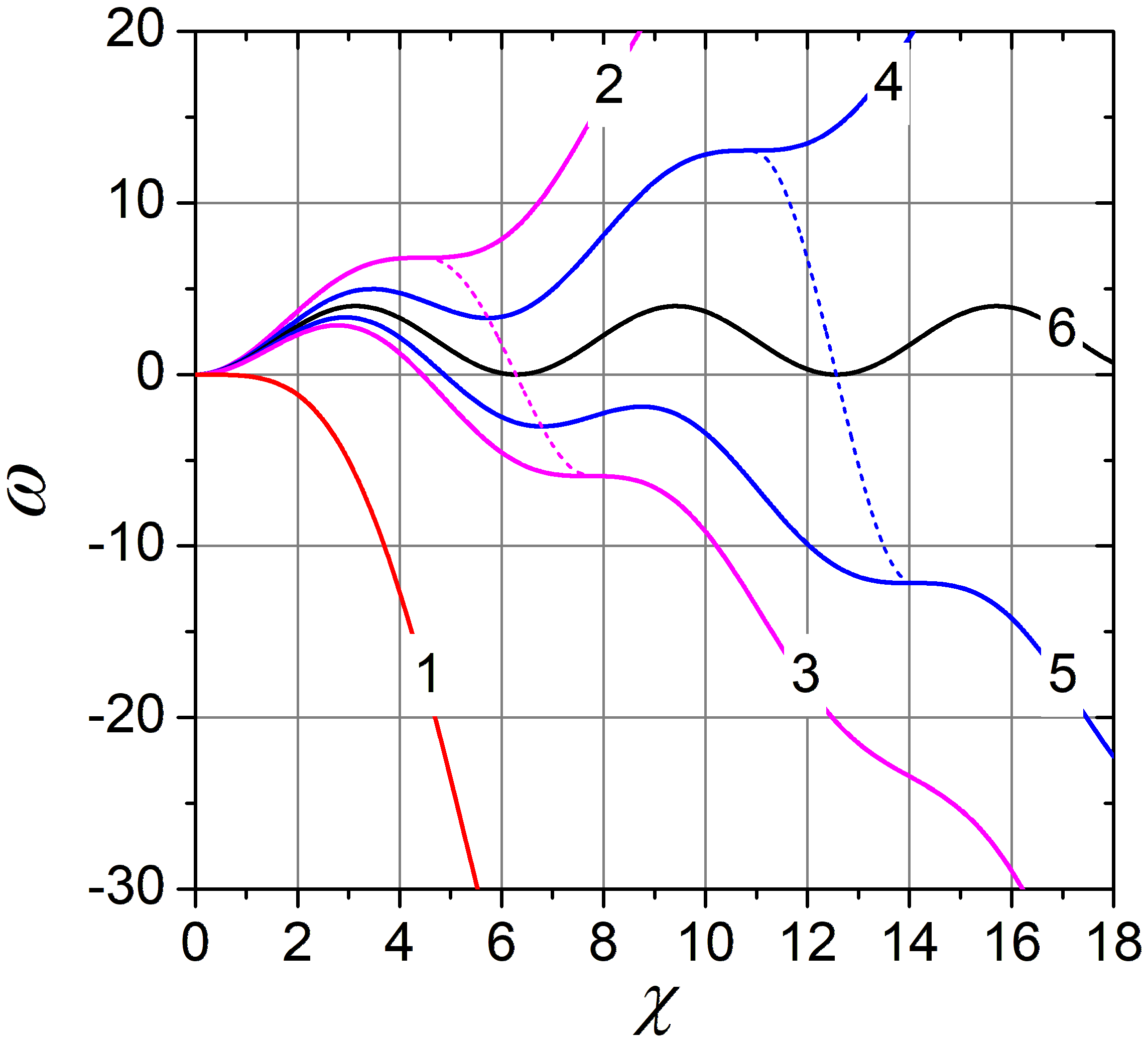} 
\vspace*{-1mm}
\caption{\label{omm}Grand canonical potential density $\omega$
as a function of order parameter $\chi$ for various $\nu$.
The curve numbers $i=\overline{1,5}$ correspond to $\nu_i=b_i$,
see (\ref{nums}), and curve 6 is for $\nu=0$. The two dotted
curves passing through the minima of $\omega$ define the ranges
of existence of phase-1 and phase-2.}
\end{figure}

Within the range of variation of the order parameter $\chi$ depicted in
Fig.~\ref{omm}, all curves $\omega(\chi)$ exhibit a minimum at $\chi=0$.
However, for $\nu>1$, the value $\chi=0$ determines unphysical maximum,
making curve~1 a boundary curve. At $\nu=0$, curve~6 possesses infinitely
many identical minima of $\omega$ at $\chi=2\pi n$ (where $n\in\mathbb{N}$),
corresponding to coexisting uniform phases.

We limit ourselves by considering the two distinct phases in curve~6: the phase-1
at $\chi=2\pi$ and the denser phase-2 at $\chi=4\pi$. These pure phases
manifest within specific intervals of variation for $\nu$ and $\chi$.
We have found that the phase-1 exists for $\nu\in[\nu_2; \nu_3]$,
while the phase-2 is present within the range $\nu\in[\nu_4; \nu_5]$.

Parameters $\nu_1$-$\nu_5$ determine the local criticality of the curves in
Fig.~\ref{omm}. Their values are established through conditions:
\begin{equation}\label{cc1}
\frac{\partial\omega}{\partial\chi}=0,\qquad
\frac{\partial^2\omega}{\partial\chi^2}=0.
\end{equation}

For further analysis, we utilize the spherical Bessel functions:
\begin{equation}
j_l(z)=z^l\left(-\frac{1}{z}\frac{\rmd}{\rmd z}\right)^l\frac{\sin{z}}{z},
\end{equation}
so that $j_l(0)=0$ for $l\geq1$.

We also need the extrema of $j_0(z)$. Since $j_0^{\prime}(z)=-j_1(z)$,
let us define the auxiliary sets of numbers $\{a_n\}$ and $\{b_n\}$,
where $n\in\mathbb{N}$, by $j_1(a_n)=0$ and $b_n\equiv j_0(a_n)$.
The first five elements of the sequences are
\begin{eqnarray}\label{nums}
&&
\begin{array}{ll}
a_1=0,&\ b_1=1;\\
a_2=4.493409,&\ b_2=-0.2172336;\\
a_3=7.7252518,&\ b_3=0.12837455;\\
a_4=10.90412,&\ b_4=-0.0913252;\\
a_5=14.06619,&\ b_5=0.070913459.
\end{array}
\end{eqnarray}

Thus, rewriting (\ref{cc1}) as
\begin{equation}
\nu=j_0(\chi),\qquad j_1(\chi)=0,
\end{equation}
one finds the set of solutions: $\chi_i=a_i$ and $\nu_i=b_i$.

\begin{figure}[h]
\centering
\includegraphics[width=4.5cm,angle=0]{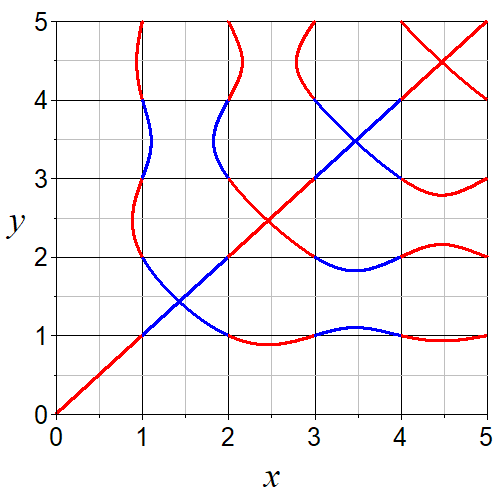} 
\caption{\label{x-y}Solution to the equation $j_0(\pi x)=j_0(\pi y)$
in the domain $x,y\in[0;5]$. Red sections correspond to
positive values, $j_0(\pi x)\geq0$, while the blue ones
are for $j_0(\pi x)<0$.}
\end{figure}

Note that the equation $\nu=j_0(\chi)$ with respect to $\chi$
can have several solutions, as shown in Fig.~\ref{x-y}.

After determining the ranges in which pure phases exist, we examine
their manifestations by turning to thermodynamic functions. The energy
density $e$ and free energy density $f$ coincide at zero temperature
to be
\begin{equation}
f=e=\omega+\nu\eta=2(1-\cos{\chi}).
\label{eps}
\end{equation}
By introducing the free energy density $f$ via the Legendre
transform of the grand potential density $\omega$, we transition from
the grand canonical ensemble (fixed $\nu$, fluctuating $\eta$) to
the canonical ensemble with a given $\eta$ (or $\chi$).

Then the chemical potential $\mu$ and the internal pressure $p$ as functions
of $\chi$ are
\begin{eqnarray}
\mu&\equiv&\frac{\partial f}{\partial\eta}=j_0(\chi),
\label{muu}\\
p&\equiv&-\frac{\partial(vf)}{\partial v}=\eta\frac{\partial f}{\partial\eta}-f
\nonumber\\
&=&\chi\sin{\chi}-2(1-\cos{\chi}).
\end{eqnarray}
Therefore, the equation $\partial\omega/\partial\chi=0$ implies
the condition $\mu(\chi)=\nu$, as before, which reflects
the equality of chemical potentials in the two ensembles used.
Then all extrema of the potential $\omega$ in Fig.~\ref{omm}
are linked by the single function $\widetilde{\omega}(\chi)=-p(\chi)$,
so that the pink and blue dotted curves are its fragments.

Note that the chemical potential $\mu$, internal pressure $p$
and internal energy density $e$ obey the Duhem--Gibbs
and Euler relations at zero temperature~\cite{LL}:
\begin{equation}
\rmd p=\eta\,\rmd\mu,\qquad
e=\mu\eta-p.
\label{DGE}
\end{equation}

Expanding $p$ for relatively small density $\eta$, the equation
of state takes on the form:
\begin{equation}\label{EoS0}
p=-\frac{\eta^2}{12}+\frac{\eta^3}{180}+O(\eta^4).
\end{equation}
The terms $-\eta^2$ and $\eta^3$ can be related to two-particle
attraction and three-particle repulsion.
The interplay of these effects may induce a first-order phase
transition, although such a transition is anticipated at
higher densities.

The configurations of axion stars with the equation of state similar to (\ref{EoS0})
were studied in detail in the work of Chavanis~\cite{Ch2018}, which predicts
the existence of a dilute and dense phase using the mass-radius
relation, remarking that analogous phases can manifest in axionlike DM.

\begin{figure}[h]
\centering
\includegraphics[width=5cm,angle=0]{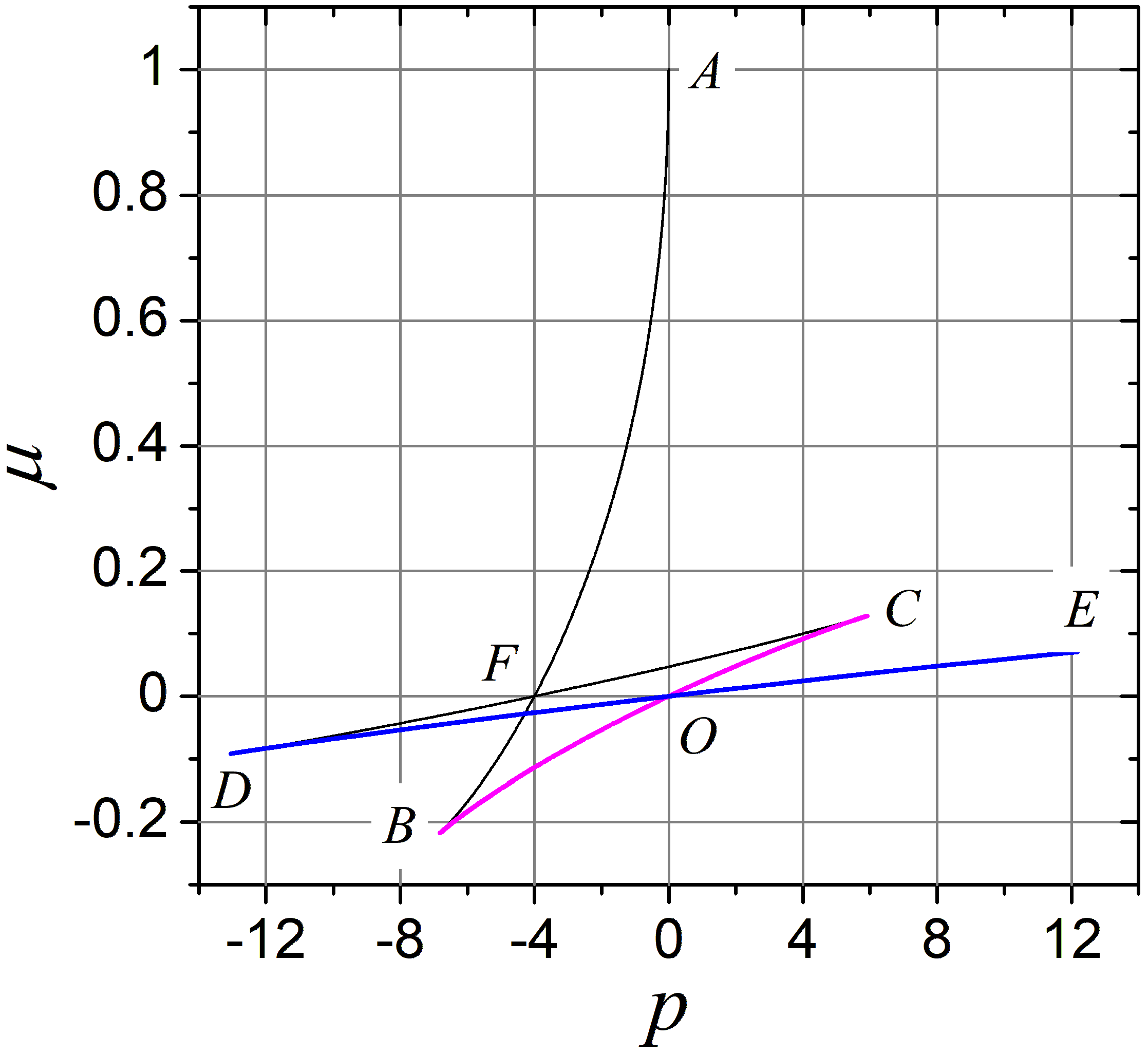} 
\caption{\label{pnu}Chemical potential $\mu$ versus internal pressure $p$,
which are parametrized by $\chi$ in the range of $\chi\in[0; 4.5\pi]$.
Segments $AB$ and $CD$
are unstable states; $BO$ and $OE$ represent stable states of phase-1
and phase-2; $OC$ and $DO$ correspond to metastable states of phase-1
and -2. The first-order phase transition between phase-1 and -2 occurs
at point $O$.}
\end{figure}

Let us analyze the $p-\mu$ phase diagram in Fig.~\ref{pnu}, which
reflects all states for $|\chi|\leq4.5\pi$. Guided by the minimum of
potential $\omega$, as shown in Fig.~\ref{omm}, we conclude that for
$\mu<0$, the phase-1 is favored over the metastable denser phase-2,
whereas for $\mu>0$, the phase-2 becomes dominant. At the same time,
Fig.~\ref{pnu} also shows {\it unstable states} $AB$ and $CD$ corresponding
to {\it local maxima} of the potential $\omega$, when
\begin{equation}
\frac{\partial\omega}{\partial\chi}=0,\qquad
\frac{\partial^2\omega}{\partial\chi^2}<0.
\end{equation}

Considering the parametric $p-\mu$ dependence for arbitrarily large $\chi$,
one readily finds that the value $\chi_c\equiv a_2$ indicates the {\it threshold}
for the emergence of stable phase in BEC DM. We stress once again that
the interval $\chi\in[0; \chi_c]$ -- depicted as segment $AB$ in Fig.~\ref{pnu}
-- represents unstable state.

Thus, the simplifications adopted in the small-$\chi$ regime restrict
the analysis to the unstable state only, thereby missing the complex -- physically
rich -- picture of phenomena.

In Fig.~\ref{pnu}, the turning points at $\chi=a_n$, where the phase pairs
converge, exhibit divergent compressibility, as determined by the expression:
\begin{equation}
-\frac{\rmd\chi}{\rmd p}=\frac{1}{\chi^2 j_1(\chi)}.
\end{equation}
This property is sometimes used to identify quantum phase transitions~\cite{NT16}.
It is also related to the (dimensionless) speed of sound, the square of which is
\begin{equation}\label{ssp}
c^2_s(\chi)=\frac{\rmd p}{\rmd\eta}=-\frac{\chi}{2}j_1(\chi).
\end{equation}
We see that $c^2_s(a_n)=0$, and $c^2_s(\chi)<0$ in
unstable states, for example, for $\chi\in[0; \chi_c]$.
Note that similar situations are known in the
laboratory~\cite{LW2008,GG2008,Dantas98}.
Besides, the $ABCDE$ diagram in Fig.~\ref{pnu} is
diffeomorphic to the {\it butterfly} bifurcation in catastrophe
theory~\cite{Arnold}.

In mechanically unstable states with $c_s^2<0$, the sign of
pressure $p$ can distinguish two contrasting patterns. 
For $p<0$ (tensile instability), the system preferentially cavitates along surfaces,
producing a foamlike structure.
In compression mode at $p>0$, the matter breaks down into discrete droplets that
may grow and coalesce.

If the admissible range of the order parameter $\chi$ does
not exceed $\chi=a_3$, implying that no phases other than phase-1
exist, then the entire segment $BC$ of phase-1 can be regarded as stable.

Without loss of generality, let us shift the potential $\omega$
by an arbitrary constant, thereby redefining its origin:
\begin{equation}
\omega=-p_0-2\cos{\chi}-\nu\chi^2.
\end{equation}
This leads to the corrected functions:
\begin{eqnarray}
&&p=p_0+\chi\sin{\chi}+2\cos{\chi},
\label{pre}\\
&&e=-p_0-2\cos{\chi},
\end{eqnarray}
while maintaining the form of $\mu=j_0(\chi)$ and (\ref{DGE}).

The value of constant $p_0$ for an inhomogeneous body in a vacuum
should be determined from the conditions $p(\chi_B)=0$ and $\mu(\chi_B)=0$
for a boundary value $\chi_B$. It follows from them that
$\chi_B=\pi k$ and $p_0=-2\cos{(\pi k)}$ for $k\in\mathbb{Z}$.
Thus, $p_0=-2$ for $k=2n$ or $p_0=2$ for $k=2n+1$, $n\in\mathbb{Z}$.

Note that the value $p_0=-2$ ensures the convergence of integrals of motion
in the spatially one-dimensional sine-Gordon theory~\cite{Man04}.
In contrast, we consider a three-dimensional model in a finite spherical volume.

{\bf Thomas--Fermi approximation.} It involves disregarding the
contribution of quantum fluctuations $\Delta_\xi\chi$ in
Eq.~(\ref{feq1}), while considering the gravitational interaction,
which leads to the set of equations:
\begin{eqnarray}
&&\hspace*{-6mm}
j_0(\chi)=\nu+\frac{A}{\xi}\int_0^\xi\upsilon(s)\,\rmd s,\quad
j_0(\chi(0))=\nu,\label{TF1}\\
&&\hspace*{-6mm}
\frac{\rmd\upsilon}{\rmd\xi}=\xi\chi^2(\xi),\hspace*{24mm}
\upsilon(0)=0,\label{TF2}
\end{eqnarray}
where the gravitational potential $\Phi(\xi)$ is identically
re-written as
\begin{equation}
\Phi(\xi)=\frac{1}{\xi}\int_0^\xi\upsilon(s)\,\rmd s,\quad
\upsilon(\xi)=\int_0^\xi\chi^2(s)\,s\,\rmd s.
\end{equation}

\begin{figure}[h]
\centering
\includegraphics[width=4.6cm,angle=0]{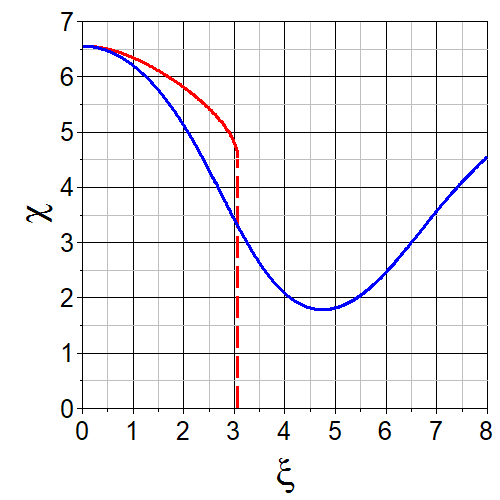} 
\vspace*{-1mm}
\caption{\label{sols}Wave functions
at $A=4.5\times10^{-3}$ and $\chi(0)=6.551147852$. Red line is
the solution of Eqs.~(\ref{TF1})-(\ref{TF2}) at
$\nu=0.04041539748$, and it terminates at $\xi_B=3.0747465$,
indicated by the vertical dashed line. Blue curve represents
the solution of Eqs.~(\ref{feq1})-(\ref{feq2})
at $\nu=0.36$.}
\end{figure}

Typical numerical solution of Eqs.~(\ref{TF1})-(\ref{TF2}) 
is shown in Fig.~\ref{sols} (red line). It describes a drop of
BEC DM in the phase-1. Physically, it is assumed that at the
origin, the dense phase is concentrated, exhibiting a predominance
of self-repulsion. Therefore, we take $2\pi<\chi(0)<2.5\pi$,
according to Fig.~\ref{pnu}, to get $j_0(\chi(0))>0$. 

To find out the reason for the discontinuity of the solution, let us
substitute the gravitational potential to obtain a closed equation
for the wave function $\chi$:
\begin{equation}\label{TFeq}
j_1(\chi)\Delta_\xi\chi
+\left(\chi^\prime\right)^2
\left[\frac{j_1(\chi)}{\chi}-j_2(\chi)\right]+A\chi^2=0.
\end{equation}
where $\chi^\prime(\xi)\equiv\partial_\xi\chi(\xi)$.

It is easy to see that the spatial development stops when
the first term vanishes due to $j_1(\chi)=0$, which happens at
$\chi=a_2$. Such $\chi$ is actually reached at $\xi_B$ in 
Fig.~\ref{sols}. At $\xi>\xi_B$ we have to put $\chi=0$,
but the gravitational potential behaves as $1/\xi$.

In principle, integration would be interrupted for any model at
$\chi(\xi_B)=a_i$ if $\chi(0)\in(a_i; a_{i+1})$, where the
five numbers of $\{a_n\}$ are shown in (\ref{nums}).

Since the nonrelativistic gravitational potential breaks the symmetry
with respect to the shift $\chi\mapsto\chi+2\pi$, let us analyze
the side effects. Note that for $V_{\rm gr}=0$ the invariance of
the functional (\ref{G1}) in terms of dimensionless variables is
ensured, in particular, by the given pseudoscalarity of $\chi$.

So, let us write the wave function as $\chi=2\pi+\widetilde{\chi}$
and consider the Poisson equation:
\begin{equation}
\Delta_\xi\Phi=\chi^2,\qquad \Phi(0)=0.
\end{equation}

We immediately see that the extracted constant contribution
$4\pi^2$ to density produces such a term in $\Phi$:
\begin{equation}
\Phi_{\rm rot}(\xi)=\frac{2\pi^2}{3}\xi^2.
\end{equation}
Its contribution to Eq.~(\ref{feq1}) may be associated with a
rotation, the role of which is already noted (see \cite{Zhang2018,Naz2020}).
If the system remains unbounded, nonrelativistic rotation with
increasing $\xi$ produces a divergent, node-free solution $\chi(\xi)$,
signaling inherent instability. However, periodic self-interaction
of limited magnitude may be insufficient to compensate for this divergence.
The finite part of such a solution to (\ref{feq1})-(\ref{feq2}),
the search of which is discussed below, is plotted in Fig.~\ref{sols}
(blue curve).
Nevertheless, similar solutions turn out to be suitable for
describing a number of observed data (see Fig.~\ref{NGC2366}).

Moreover, we refer to the model from \cite{GKN20},
which suggests to separate the approaches to describing the core and
the tail of the DM halo due to different role of self-interaction in
relatively dense and rarefied regions.

\subsection{\label{S2C}The excitation spectrum and instabilities}

Given the quantum-fluid nature of BEC DM, let us analyze the spectrum
of its small excitations. We reformulate Eq.~(\ref{feq1}) into the
non-stationary Gross--Pitaevskii equation for the complex-valued wave
function $X$, which depends on dimensionless time $t$, as
\begin{eqnarray}
&&
\left(-\Delta_{\xi}+W\right)X=\rmi\partial_tX,
\label{feq3}\\
&&
W(\xi,t)=A{\cal V}(\xi,t)+j_0(|X(\xi,t)|),
\label{feq4}\\
&&
{\cal V}(\xi,t)=-\frac{1}{\xi}\int_0^\xi |X(s,t)|^2s^2\,\rmd s
\nonumber\\
&&\hspace*{14mm}
-\int_{\xi}^{\xi_B}|X(s,t)|^2 s\,\rmd s,
\label{feq5}
\end{eqnarray}
where $\rmi=\sqrt{-1}$.

To derive both the continuity equation and the Euler hydrodynamic equation,
we substitute $X=\chi\exp{(\rmi\theta)}$,
where $\chi$ and $\theta$ are real functions of space and time.

Substituting the wave function $X$ into (\ref{feq3}) and separating the imaginary
and real parts, we obtain the desirable equations:
\begin{eqnarray}
&&\hspace*{-2mm}
\partial_t\eta+{\bg\nabla}(\eta{\bf v})=0,
\label{conr}\\
&&\hspace*{-2mm}
\partial_t{\bf v}+{\bg\nabla}\left(\frac{{\bf v}^2}{2}+2W-\frac{2}{\sqrt{\eta}}\Delta_\xi\sqrt{\eta}\right)=0,
\label{Ehr}
\end{eqnarray}
where the velocity vector ${\bf v}\equiv2{\bg\nabla}\theta$ is expressed
via the gradient operator ${\bg\nabla}$; and $\eta=\chi^2$ is the particle
number density, as before.

Consider a thin spherical shell defined by the radial interval from $\xi$
to $\xi+\delta\xi$ for small $\delta\xi$. Within this shell, we admit
small perturbations $\delta\eta$ in the particle density and $\delta{\bf v}$
in the collective velocity, added to the fixed background values
$\eta_0$ and ${\bf v}_0=0$ with the subscript ``0''.
Further, we use Eqs.~(\ref{conr})-(\ref{Ehr}), linearized in
$\delta\eta$ and $\delta{\bf v}$.

Differentiating the continuity equation and substituting the Euler equation
to eliminate $\delta{\bf v}$, we arrive at
\begin{eqnarray}
0&=&\partial_t\left[\partial_t\delta\eta+\eta_0{\bg\nabla}\delta{\bf v}\right]
\nonumber\\
&=&\partial^2_t\delta\eta-\eta_0\Delta_\xi\left(2W-\frac{1}{\eta_0}\Delta_\xi\delta\eta\right).
\end{eqnarray}
This equation admits that the background value $\eta_0$ evolves more slowly
over time and space than the perturbation $\delta\eta$. In the absence of
gravitation, the homogeneity of matter is allowed, that is, when $\eta_0$
is constant.

Since the interaction $W$ comprises both the gravitational term, governed
by the parameter $A$, and the self-interaction, we expand it as a series,
retaining only the linear corrections in $\delta\eta$:
\begin{equation}
W=W_0+A\Delta^{-1}_\xi\delta\eta+c^2_s(\chi_0)\frac{\delta\eta}{\eta_0},
\end{equation}
by using the formula for the sound speed $c_s$, see (\ref{ssp}).

Assuming that inside the thin shell the gravitational potential ${\cal V}_0$
can be replaced by the Newtonian $\sim1/\xi$, and the self-interaction
$j_0(\chi_0)$ remains constant, we derive the equation:
\begin{equation}
\partial_t^2\delta\eta-2c^2_s(\chi_0)\Delta_\xi\delta\eta-2A\eta_0\delta\eta+\Delta^2_\xi\delta\eta=0.
\end{equation}

It is worth noting other arguments in \cite{Ki03,FHWM} that exclude
the influence of ${\cal V}_0$ on the evolution of excitations.
 
Setting
\begin{equation}
\delta\eta=\chi_0(\delta\chi_k+\delta\overline{\chi}_k),
\end{equation}
where $\delta\overline{\chi}_k$ means the complex conjugate of $\delta\chi_k$,
we define the spherically symmetric form of excitation as
\begin{eqnarray}
&&\delta\chi_k=\chi_{+k}f_k(\xi,t)+\chi_{-k}\overline{f}_k(\xi,t),
\label{flu}\\
&&
f_k(\xi,t)=\frac{1}{\xi}\exp{(\rmi k\xi-\rmi\omega_k t)},
\end{eqnarray}
where $\chi_{\pm k}$ are real amplitudes of incident and outgoing
spherical waves; $\omega_k$ is their frequency indexed by
the wave number $k$.

Substituting, we obtain the dispersion relation:
\begin{equation}\label{disr}
\omega^{(\pm)}_k=\pm\sqrt{k^4+2k^2c^2_s(\chi_0)-2A\eta_0}.
\end{equation}
Its dimensional form looks like
\begin{equation}\label{disr2}
\frac{E^{(\pm)}_k}{\hbar}=\pm\sqrt{\frac{\hbar^2}{4m^2}q^4+q^2v^2_s-4\pi G\rho},
\end{equation}
where the wave vector squared $q^2=k^2/r_0^2$,
the mass density $\rho=\rho_0\eta_0$, and the squared sound
speed
\begin{equation}
v^2_s=\frac{\hbar^2}{2m^2r_0^2}c^2_s.
\end{equation}

This dispersion relation is well known~\cite{Ch2011,Ber25,H2019,Ch20}. And
the models differ in the properties of the sound speed~$v_s$.

Let us first analyze the case without accounting for gravitation,
when $A=0$ and the dispersion relation is reduced to the form:
\begin{equation}\label{exc}
\omega_k^{(\pm)}=\pm\sqrt{k^4-k^2\chi_0j_1(\chi_0)}.
\end{equation}

We immediately see that at the turning points in Fig.~\ref{pnu},
i.e. $\chi_0=a_n$ from the sequence $\{a_n\}$ in (\ref{nums}),
the sound speed vanishes, $c^2_s(a_n)=0$, and the free-particle
spectrum, $\omega_k^{(\pm)}=\pm k^2$, is recovered.

Since $j_1(\chi_0)$ remains positive for $\chi_0\in(a_{2n-1}; a_{2n})$,
$n\in\mathbb{N}$, the excitation modes with $k^2<\chi_0j_1(\chi_0)$
become unstable due to the imaginary $\omega_k^{(\pm)}$.
These instability conditions are consistent with those predicted
thermodynamically, in particular, for the segments $AB$ and $CD$
in Fig.~\ref{pnu}.
On the other hand, the physically constructive phase-1 and phase-2 correspond
to $\chi_0$ in the intervals $(a_2; a_3)$ and $(a_4; a_5)$, where
$j_1(\chi_0)<0$ and $c^2_s>0$. Thereby, Eq.~(\ref{exc}) also confirms
the sustainability of these states.
In this case, the Bogolyubov spectrum is obtained,
similar to that in superfluidity models~\cite{Ber25}.

Therefore, stable structures of axionlike DM are feasible at
$\chi\in(a_{2n}; a_{2n+1})$, i.e. at $\chi>\chi_c$. In this range,
a nonperturbative approach becomes necessary.

For a nonzero parameter $A\ll1$, to overcome the gravitational
Jeans instability at $k\to0$ in (\ref{disr}), it is necessary
to introduce a lower bound $k_J$ for wave numbers $k$.
However, due to the substantially inhomogeneous DM distribution
and the need for a nonperturbative approach, the limiting size
of a stable DM cloud cannot be reliably determined based on $k_J$.
Therefore, studies using time-dependent numerical analysis are needed
to refine Jeans limit in size and mass. For this reason, we adopt
the known limits for cold boson stars~\cite{BGZ84,UTB02} and
conservatively estimate the maximum stable mass as $M\sim0.6\hbar c/(Gm)$.

\section{\label{S3}Examining the model in\\the context of NGC 2366}

\subsection{\label{S3A}Solving the field equations and\\fitting the rotation curve}

Due to the nonlinearity of Eq.~(\ref{feq1}), the initial conditions
of the spatial evolution of the field $\chi$ cannot be given arbitrarily,
but must be determined. By requiring the absence of cuspidity at the origin, i.e.
$\chi(0)<\infty$, $\chi^\prime(0)=0$, and $\chi^{\prime\prime}(0)<0$ for given
$A$ and $\nu$, the finite initial value $\chi(0)=z$ should satisfy the equation:
\begin{equation}\label{icon}
Az^2+(\nu-j_0(z))(\nu-\cos{z})=0.
\end{equation}
This is derived by substituting the expansion $\chi(\xi)=\chi(0)+\chi^{\prime\prime}(0)\xi^2/2$
for $\xi\to0$ into (\ref{feq1})-(\ref{feq2})  and by finding $\chi^{\prime\prime}(0)$ \cite{GN23}.
Note that the condition $\chi^{\prime\prime}(0)<0$ is equivalent to
requiring $j_0(z)<\nu<\cos{z}$ for positive~$\nu$.

\begin{figure}[h]
\centering
\includegraphics[width=5cm,angle=0]{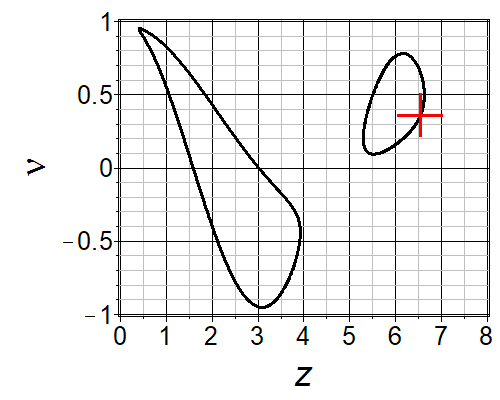} 
\vspace*{-1mm}
\caption{\label{ics}Black curves represent the solutions of Eq.~(\ref{icon})
for $A=4.5\times10^{-3}$. The red cross marks the initial condition
for the blue curve in Fig.~\ref{sols}.}
\end{figure}

The form of Eq.~(\ref{icon}) indicates that its solution $z$ exists
for small $A$. For this reason, we take $A$ to be of the order of
$10^{-3}$. Physically, this means that gravitational interaction is
significantly weaker than the self-interaction between particles.
However, we immediately encounter the problem of the solution ambiguity
as we see in Fig.~\ref{ics}. Moreover, the number of ``islands'',
as in Fig.~\ref{ics}, increases with decreasing $A$.
As we shall see, this ambiguity leads to solutions with different
thermodynamic properties.

The absence of any solution $z$ for a given pair $(A,\nu)$ means that
$\chi(\xi)=0$ everywhere. Except for large $A$, this happens for $\nu$
outside the range $[\nu_{\min};\nu_{\max}]$ for each ``island'', where
$\nu_{\max}(A)$ and $\nu_{\max}(A)$ are also found numerically from (\ref{icon}).
For $\nu_{\min}<\nu<\nu_{\max}$, two branches of $\chi_0(\nu)$ can occur,
which indicate the existence of two regimes  and possibility of a
first-order phase transition. 

Before theoretically studying the effects in DM, we test the model
ability to describe observables. It is natural to focus on the rotation
curves (RCs) of galaxies. BEC DM models have already been successfully
employed to reproduce RCs for galaxies from
the SPARK database~\cite{Harko2011,Craciun2019,Kun2020}.

The RC of a DM-dominated (dwarf) galaxy can be derived
in the parametric form:
\begin{eqnarray}
&&\hspace*{-8mm}
V(r)=\{(r(\xi),\,V(\xi))\,|\,r=r_0\xi,\,V=V_0v(\xi)\},
\\
&&\hspace*{-8mm}
v(\xi)=\sqrt{\frac{n(\xi)}{\xi}},\quad
n(\xi)=\int_0^{\xi}\chi^2(s) s^2\rmd s,
\label{vn}
\end{eqnarray}
where $r_0$ and $V_0$ are dimensional scales; $n(\xi)$ defines
the number (or mass) of particles inside a sphere of radius $\xi$.

Having obtained the optimal parameters $A$, $V_0$, and $r_0$,
we can extract the dark axion mass $m$, the mass density
scale $\rho_0$ defining the central density $\rho_c=\rho_0\chi^2(0)$,
and the axion decay constant $f_\text{a}$ as
\begin{eqnarray}
&&m=\sqrt{\frac{A}{2}}\frac{\hbar}{V_0r_0},\quad
\rho_0=\frac{V_0^2}{4\pi G r_0^2},\quad
\nonumber\\
&&
f_\text{a}=\frac{m_{\text{P}}V_0^2}{\sqrt{2\pi A}},
\end{eqnarray}
where $m_{\text{P}}\equiv\sqrt{\hbar c/G}=2.176434\times10^{-8}$~kg.

To facilitate estimation, we provide the formulas:
\begin{eqnarray}
m&=&0.4287\times10^{-22}~\text{eV}~c^{-2}~\left[\frac{A}{10^{-3}}\right]^{1/2}
\nonumber\\
&&\times\left[\frac{r_0}{1~\text{kpc}}\right]^{-1}
\left[\frac{V_0}{1~\text{km}/\text{s}}\right]^{-1},
\nonumber\\
\rho_0&=&0.1252\times10^{-23}~\text{kg}~\text{m}^{-3}
~\left[\frac{r_0}{1~\text{kpc}}\right]^{-2}
\nonumber\\
&&
\times\left[\frac{V_0}{1~\text{km}/\text{s}}\right]^{2},
\nonumber\\
f_\text{a}&=&0.1714\times10^{19}~\text{eV}~\left[\frac{A}{10^{-3}}\right]^{-1/2}
\left[\frac{V_0}{1~\text{km}/\text{s}}\right]^{2}.
\nonumber
\end{eqnarray}

\begin{figure}[h]
\centering
\includegraphics[width=6cm,angle=0]{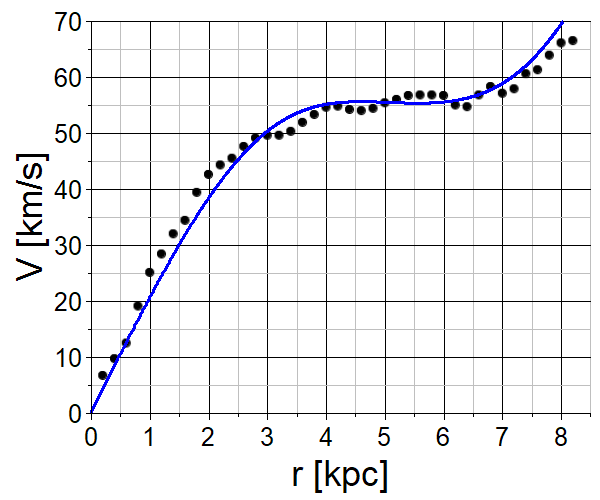} 
\caption{\label{NGC2366}Rotation curve for NGC~2366. Black dots
corresponds to the data from \cite{Oh08}. The solid curve follows
from the blue solution in Fig.~\ref{sols}.}
\end{figure}

It turns out that the blue curve in Fig.~\ref{sols} can be used
to reproduce the {\it adapted} RC of {\it irregular} dwarf galaxy
NGC~2366 in Fig.~\ref{NGC2366} for $r_0=1.18$~kpc and $V_0=6.58$~km/s.
Although we have achieved a good fit, a realistic picture requires
at least taking into account the small contribution of the baryon
distribution, which is here absorbed by the effective chemical
potential $\nu$.

Although the rescaled RC in Fig.~\ref{NGC2366}
superficially resembles, for example, the Milky Way's DM contribution
in \cite{Zhang2018}, we stay with the selected dwarf galaxy, whose internal
processes are important for our study.

The result of fitting the RC leads to the following values of DM
characteristics: the dark axion mass $m=0.1171\times10^{-22}~\text{eV}c^{-2}$,
the decay constant $f_\text{a}=3.4978\times10^{19}$~eV, and
the central mass density $\rho_c=0.1671\times10^{-20}~\text{kg}/\text{m}^3$.
Although the values of $m$ and $\rho_c$ appear to be an order of
magnitude smaller than expected, they are not surprising given
the known results~\cite{KMT}.

Note that various methods for estimating dark‐boson masses yield a wide,
often order‐of‐magnitude, spread in their values~\cite{Zim25}.
The anisotropy of the CMB requires
$m>10^{-24}~\text{eV}c^{-2}$~\cite{Ev1}, and taking into account
weak gravitational lensing gives $m>10^{-23}~\text{eV}c^{-2}$~\cite{Ev2}.
The Lyman-$\alpha$ forest simulations raise the lower bound to
$m>10^{-20}~\text{eV}c^{-2}$~\cite{Ev3}.
It seems that each model determines the mass that best suits
a particular regime.

Note that single-parameter fuzzy-DM models face the {\it Catch-22}
problem~\cite{Marsh2014}: the boson mass must be light enough to generate kpc-scale cores in
dwarf-galaxy halos, requiring $m\simeq10^{-22}\,\mathrm{eV}/c^2$, yet heavy
enough to preserve small-scale power in the Lyman-$\alpha$ forest. Introducing
at least one additional parameter, which implies a self-interaction presence,
can break this degeneracy and reconcile core formation with Lyman-$\alpha$
constraints. In this context, it is worth noting the detailed study of SIDM 
models, when the first terms of the Taylor series for potential~(\ref{SI1})
are taken into account at a relatively low density~\cite{Ch2011,Ch2016,Mocz2023,Paint2024}.

Found boson mass $m$ allows us to roughly estimate the limiting mass of
gravitationally stable DM sphere,
$M_c\sim0.6\hbar c/(Gm)\simeq6.84\times10^{13}M_\odot$,
while the computed DM halo mass is about $10^{10}M_\odot< M_c$,
where $M_\odot$ is the solar mass. This indicates
the gravitational stability of the DM in NGC~2366.

Another fitting of the RC of NGC~2366 in the BEC DM concept is shown
in \cite{Harko2011}. There, the positive $s$-scattering length $a_s$
and wave number $k$, $k^2=Gm^3/(\hbar^2a_s)$, are used. In contrast,
if the pairwise interaction $\psi^4$ is naively isolated from
potential (\ref{SI1}) in the vicinity of $\psi=0$, which corresponds
to an unstable state, we obtain a negative, usually underestimated,
scattering length from the relation:
\begin{equation}
-\frac{Uv}{24}=\frac{2\pi a_s\hbar^2}{m}.
\end{equation}
Using it, we can formally write that
\begin{equation}
A=\frac{Gm^3}{\hbar^2|a_s|}\frac{r_0^2}{6},
\end{equation}
to relate the dimensionless $A$ with $k$ in \cite{Harko2011}.

On the other hand, by expanding potential (\ref{SI1}) in a series in
the vicinity of some $\psi$ belonging to a stable state, for example,
phase-1, one may obtain a positive scattering length $a_s>0$, as it
is expected.

\begin{figure}[h]
\centering
\includegraphics[width=7.7cm,angle=0]{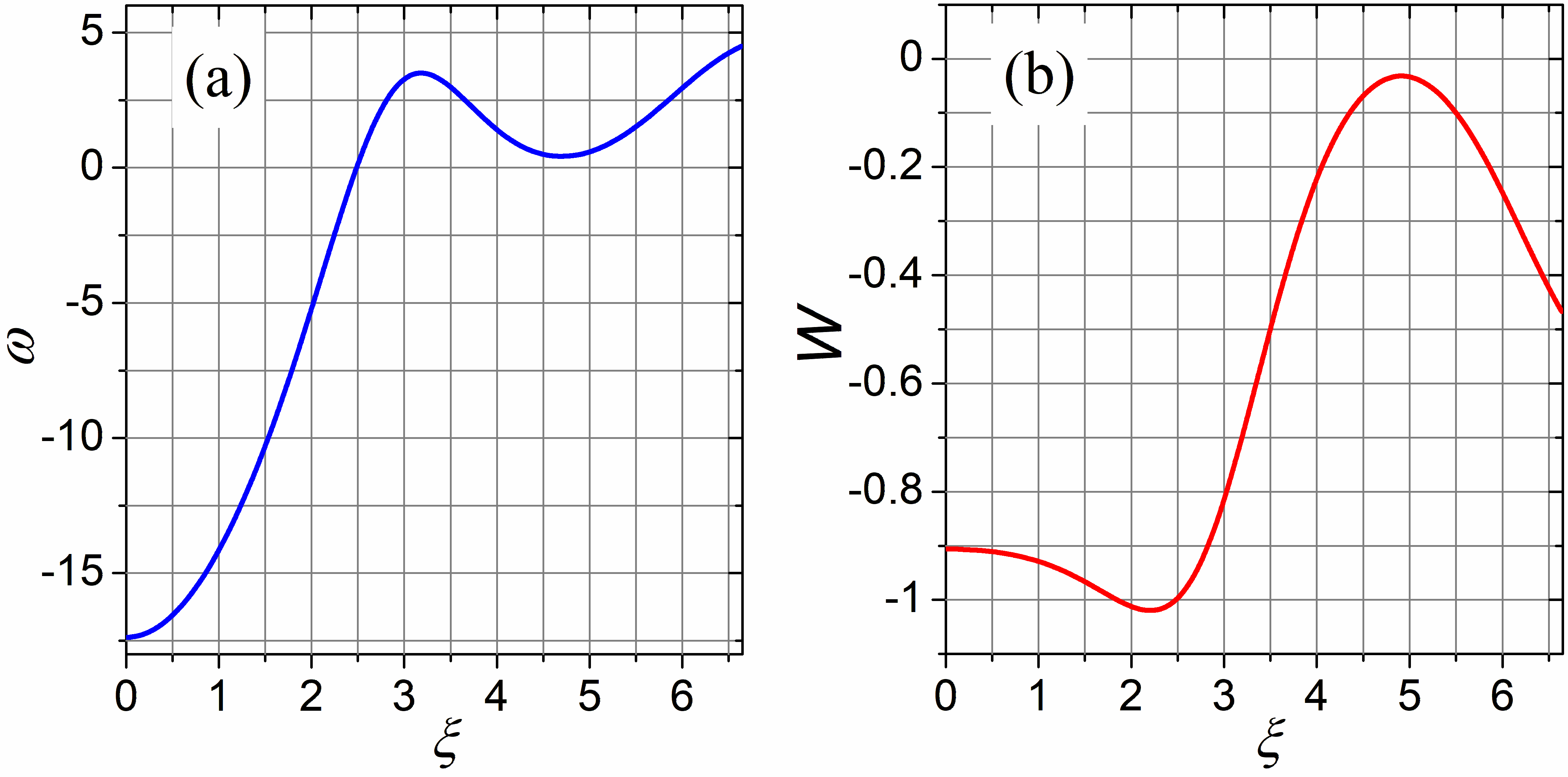} 
\caption{\label{pots}Spatial dependence of the grand potential density (a)
and the quantum potential (b) for the blue solution in Fig.~\ref{sols}.}
\end{figure}

It is useful to turn to the (shifted) grand canonical potential density $\omega$
and the potential $W$:
\begin{eqnarray}
&&\omega=(\partial_\xi\chi)^2-\nu\chi^2+A\Phi\chi^2-2\cos{\chi},\\
&&W(\xi)=A\Phi_0+A\Phi(\xi)+j_0(\chi(\xi)).
\end{eqnarray}
The latter is involved in the Schr\"odinger-type equation
$[-\Delta_{\xi}+W(\xi)]\chi=\varepsilon\chi$, where 
$\varepsilon=1+2u$, and $\varepsilon\simeq-0.5867$ for
the DM in NGC~2366.

\begin{figure}[h]
\centering
\includegraphics[width=4cm,angle=0]{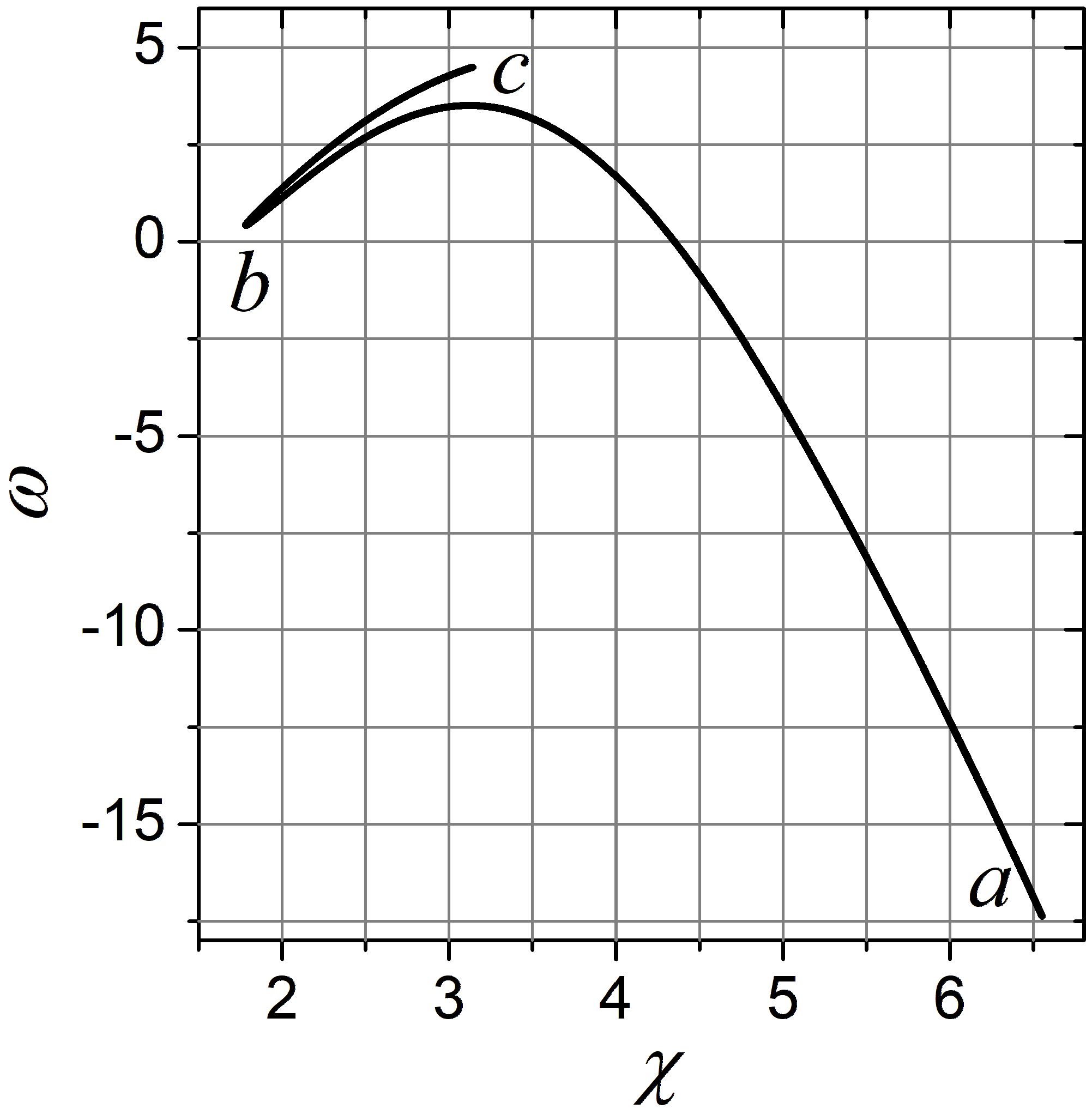} 
\caption{\label{omx}Dependence of the grand potential density $\omega$
on the order parameter $\chi$ determined by the blue solution in Fig.~\ref{sols}.}
\end{figure}

For the parameters and solution $\chi(\xi)$ leading to the RC of NGC~2366,
the profiles of these potentials are shown in Fig.~\ref{pots}. Both indicate
a potential well in the core region and have an unstable tail.

However, we wish to highlight a remarkable property of $\omega$ within
the interval \mbox{$2.5<\xi<6.3$}, corresponding to the rarefied unstable
phase. Specifically, within this band, the potential $\omega$ can take on the
same value twice or even three times. This phenomenon is naturally linked
to the repeated values of $\chi(\xi)$ in Fig.~\ref{sols}.
The ambiguity is also visible in Fig.~\ref{omx}, where the curve segment $bc$
covers part of $ab$. It is natural to assume that, for certain $\chi$,
the potential $\omega$ becomes ambiguous due to varying influences.

Thereby, we have identified transient processes
occurring in the tail of the DM distribution in NGC~2366.
Further, we explore these processes in more detail,
focusing on solutions $\chi(\xi)$ similar to the blue one shown
in Fig.~\ref{sols}, and using a statistical approach.

\subsection{\label{S3B}Analyzing the macroscopic state of BEC DM}

Having extracted the model parameters from the galaxy’s RC fit
(e.g., NGC 2366), we then apply statistical analysis to probe the
thermodynamic properties of BEC DM. Specifically, we focus on identifying
the macroscopic state of DM in the galaxy. It depends on the fractions
of the dense and rarefied phases of DM, and their change may be
a phase transition. By mapping all configurations onto a diagram,
the state of the galaxy would be represented by a single point there.

When analyzing the state of inhomogeneous matter, it is necessary
to constrain its volume by imposing boundary conditions. 
We define the boundary radius $\xi_B$ of a self-gravitating sphere in a vacuum
by the condition when the local pressure (\ref{pre}) continuously
reaches zero, $p(\chi(\xi_B))=0$. In the case of NGC 2366, $\chi(\xi)$ never
crosses zero, so we impose $\chi(\xi_B)=\pi$, which produces
a discontinuous drop in density at the surface. Under the same condition
the chemical potential also vanishes, $\mu(\chi(\xi_B))=0$. A similar sharp
boundary appears in compact, solid objects (for example, neutron stars), where
the interior matches directly onto vacuum, if there is no (thin) atmosphere of
evaporating or falling particles.

\begin{figure}[h]
\centering
\includegraphics[width=4.6cm,angle=0]{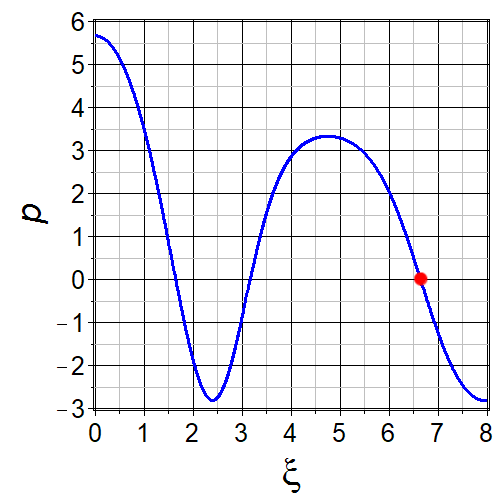} 
\caption{\label{press}Spatial dependence of local pressure
$p$ created by the blue solution $\chi(\xi)$ in Fig.~\ref{sols}
according to (\ref{pre2}). The red dot marks the boundary
of the system.}
\end{figure}

Based on the analysis of Figs.~\ref{sols}, \ref{NGC2366} and \ref{press},
we choose $\xi_B\simeq6.64$ to ensure $\chi(\xi_B)=\pi$ and to cover almost
the entire visible region of NGC 2366, which is bounded by $\xi\simeq6.95$
in dimensionless units. At the same time, the presence of 
the cold DM~\cite{CDM,NFW} beyond the galaxy is admitted, which is a common
assumption in halo models.

When generating a set of solutions $\chi(\xi)$ for different values of
$A$ and $\nu$ to support statistical analysis, we ensure that their
shapes are similar to the reference solution shown in Fig.~\ref{sols}.
To achieve this correspondence, varying $\nu$ at fixed $A$, we choose
initial conditions from an ``island'' similar to that in Fig.~\ref{ics}
and impose certain conditions:
\begin{equation}\label{bc2}
\chi^\prime(\xi_B)>0,\qquad \chi(\xi_B)=\pi,
\end{equation}
which should hold simultaneously and define $\xi_B$.

Each wave function $\chi(\xi)$ in this set defines the chemical potential
$\mu$, internal pressure $p$, and internal energy density $e$
according to
\begin{eqnarray}
&&\mu=j_0(\chi),\\
&&p=\chi\sin{\chi}+2(1+\cos{\chi}),
\label{pre2}\\
&&e=-2(1+\cos{\chi}),
\end{eqnarray}
which are connected by the Duhem--Gibbs and Euler relations (\ref{DGE}).
Thus, the $p-\mu$ diagram here is given by the diagram in Fig.~\ref{pnu}
by shifting the origin.

On the other hand, Eq.~(\ref{feq1}) requires that the chemical potential
$\mu$ be the sum of the fields:
\begin{equation}
\mu=h-A\Phi,\quad
h\equiv\nu+\frac{1}{\chi}\,\Delta_\xi\chi,
\label{mu2}
\end{equation}
where $h$ describes the long-wave quantum fluctuations~\cite{LL}.

The mean internal pressure induced by the axionlike self-interaction
is defined as
\begin{equation}
P=\frac{3}{\xi^3_B}\int_0^{\xi_B}p(\xi)\xi^2\,\rmd\xi,
\label{mpress}
\end{equation}
where $p(\xi)=p(\chi(\xi))$, substituting the Eq.~(\ref{pre2}).

Having designated the value of the wave function at the boundary as
$\chi_B\equiv\chi(\xi_B)$, and its derivative as
$\chi^\prime_B\equiv\chi^\prime(\xi_B)$, we proceed to the analysis
of Eq.~(\ref{mpress}). Integrating by parts the trigonometric
expressions, we arrive at
\begin{eqnarray}
P&=&-\chi_B\sin{\chi_B}+\frac{3}{\xi^3_B}\int_0^{\xi_B}\chi^2(\xi) \mu(\xi) \xi^2\,\rmd\xi
\nonumber\\
&&+\frac{2}{\xi^3_B}\int_0^{\xi_B}\chi^\prime(\xi) \chi(\xi) \mu(\xi) \xi^3\,\rmd\xi.
\label{mpr}
\end{eqnarray}
The first term vanishes at $\chi_B=\pi$, but it may be needed
in the general case.

Note that the functional (\ref{feq3}) is then given by
\begin{equation}
\frac{\Gamma}{\Gamma_0}=\frac{\xi_B^3}{3}\left(4-P+\frac{3}{\xi_B}\chi_B \chi^\prime_B\right).
\end{equation}
Additionally, the mean pressure is evaluated in the units
\begin{equation}
P_0\equiv\frac{\Gamma_0}{4\pi r_0^3}=\frac{\varepsilon_0}{v}=\frac{f_{\text{a}}^2}{2\hbar c r_0^2},
\end{equation}
yielding the value $P_0=2.3383~\text{eV}\,\text{cm}^{-3}=3.7464\times10^{-13}~\text{Pa}$
for the DM in NGC~2366.

Using the Eq.~(\ref{mpr}), we can establish a relationship between
the mean internal pressure and the partial pressures arising from
quantum fluctuations and self-gravity by inserting the chemical
potential from Eq.~(\ref{mu2}). Integrating, we obtain
\begin{eqnarray}
\hspace*{-2mm}
P&=&-\chi_B\sin{\chi_B}+\Pi_{\rm q}+A\Pi_{\rm gr},
\label{mpress2}\\
\hspace*{-2mm}
\Pi_{\rm q}&=&\frac{3}{\xi_B}\chi_B \chi^\prime_B+(\chi^\prime_B)^2
+\nu\chi^2_B
\nonumber\\
&&
-\frac{2}{\xi^3_B}\int_0^{\xi_B}(\chi^\prime(\xi))^2\xi^2\,\rmd\xi,
\label{qpre}\\
\hspace*{-2mm}
\Pi_{\rm gr}&=&-\Phi(\xi_B)\chi^2_B
+\frac{1}{\xi^3_B}\int_0^{\xi_B}\frac{n(\xi)\, \rmd n(\xi)}{\xi},
\label{gpre}
\end{eqnarray}
where $n(\xi)$ is defined in (\ref{vn}).
Since $\Pi_{\rm gr}<0$, one gets that $\Pi_{\rm q}>P>0$.

Thus, we extract the contribution $\Pi_{\rm q}$ of quantum
fluctuations by setting $A=0$ in (\ref{mpress2}) for a moment
and thereby dropping the self-gravity term $\Pi_{\rm gr}$. With
increasing $\Pi_{\rm q}$, i.e. with enhancing fluctuations, it
is expected an increase in the system size $\xi_B$
with a decrease in the particle number density
\begin{equation}
\sigma=\frac{3}{\xi^3_B}N;\quad
N=n(\xi_B),
\end{equation}
where $N$ defines the total number of particles.

At this stage, our goal is to determine under what conditions
such characteristics continuously change and when the phase
transition occurs. At zero temperature, this can be determined
by varying the gravitational parameter $A$.

\begin{figure}[h]
\centering
\includegraphics[width=7.7cm,angle=0]{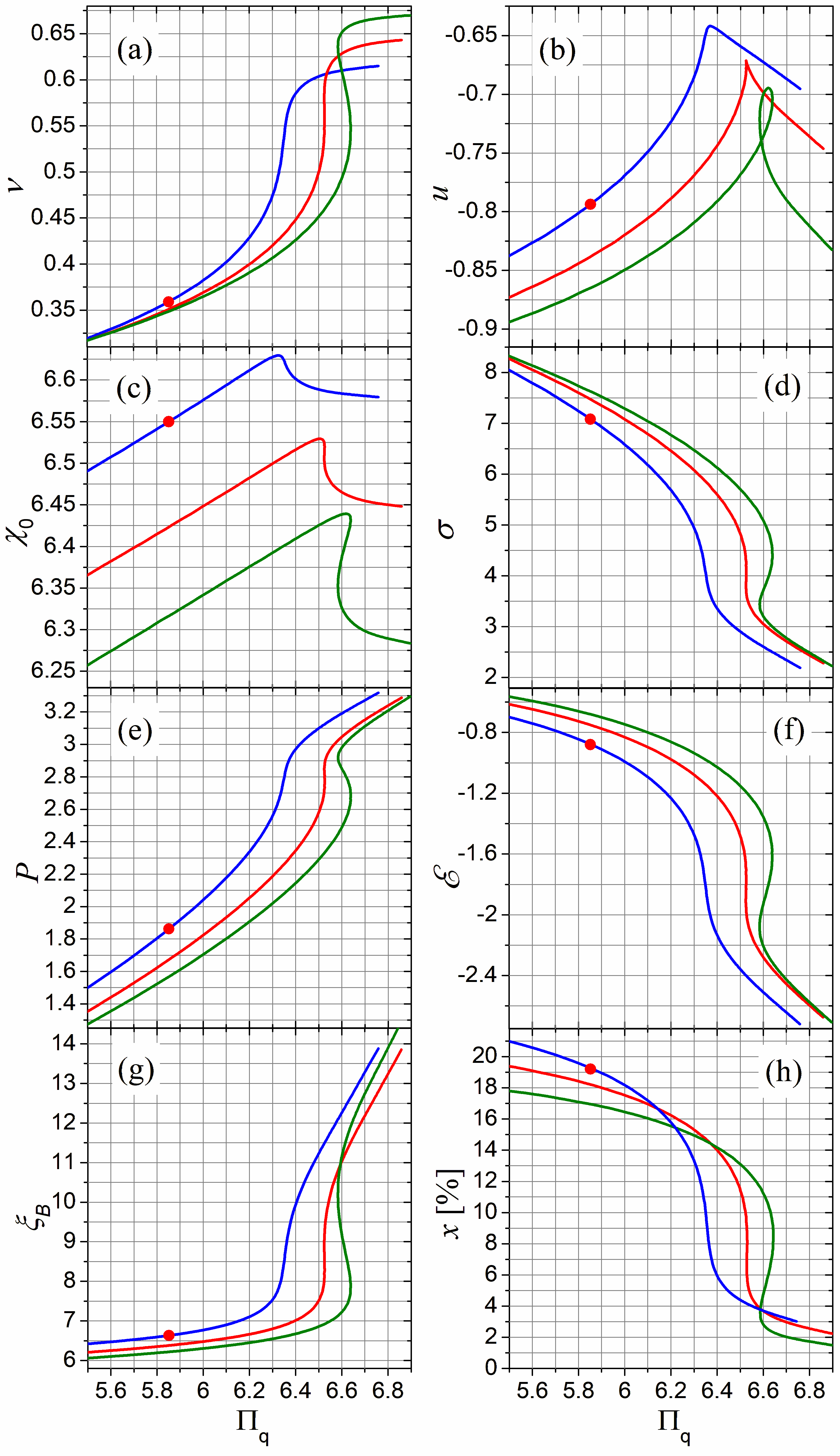} 
\caption{\label{phas}The BEC DM characteristics with changing
pressure of quantum fluctuations $\Pi_{\rm q}$ at $A=4.5\times10^{-3}$
(blue curves), $A=5.1\times10^{-3}$ (red curves), $A=5.6\times10^{-3}$
(green curves). The red dot corresponds to the solution in
Fig.~\ref{sols} applied to describe RC of NGC 2366 in
Fig.~\ref{NGC2366}.}
\end{figure}

The numerical results in Fig.~\ref{phas} indicate that DM configurations
with denser content {\it continuously} transform into a state
with less dense content (and vice versa) when $A<5.1\times10^{-3}$,
while the first-order phase transition occurs at $A>5.1\times10^{-3}$.
This implies that the value of $A=5.1\times10^{-3}$ is close to critical.

Indeed, the backbendings of the green curves for $A=5.6\times10^{-3}$
within the interval of $\Pi_{\rm q}\in(6.58;\,6.64)$ indicate the
presence of the metastable states during the phase transition.
Physically, a discontinuity in properties arises in this interval,
the position of which (i.e. value of $\Pi_{\rm q}$) is
determined by Maxwell's construction. 

Although the basic macroscopic functions at $A=5.6\times10^{-3}$,
including the mean internal pressure $P$, are single-valued functions
of $\nu$, the function $\Pi_{\rm q}$ is not.
For example, the value $\Pi_{\rm q}=6.60$ occurs three times:
at $\nu_1\simeq0.493$, $\nu_2\simeq0.611$, and $\nu_3\simeq0.651$,
see Fig.~\ref{phas}(a). Then the pressure jump in Fig.~\ref{phas}(e)
during the phase transition is equal to
the difference $\Delta P\equiv P(\nu_3)-P(\nu_1)=A(\Pi_{\rm gr}(\nu_3)
-\Pi_{\rm gr}(\nu_1))$ according to (\ref{mpress2}) for $\chi_B=\pi$.
The positivity of $\Delta P\simeq0.492$ indicates a weakening of
the gravitational interaction. The jumps of other functions should
then be expressed through $\Delta P$, which is precisely determined
by applying Maxwell's rule.

Note that this property of $\Pi_{\rm q}$ arises due to the long tail
of the rarefied unstable phase, as in Fig.~\ref{sols}, which is defined
by $\chi<\chi_c$.

Denoting $\chi_0\equiv\chi(0)$ and expanding the set of previously
introduced characteristics, we include the mean energy density
${\cal E}$ in Fig.~\ref{phas}(f):
\begin{equation}
{\cal E}=\frac{3}{\xi^3_B}\int_0^{\xi_B}e(\xi)\xi^2\rmd\xi,
\end{equation}
and the DM phase-1 fraction in Fig.~\ref{phas}(h):
\begin{equation}
x=\frac{100\%}{N} \int_0^{\xi_B}\theta(\chi(\xi)-\chi_c)\chi^2(\xi)\xi^2\rmd\xi,
\end{equation}
where $\theta(z)$ is the Heaviside function.

Thus, an increase in quantum fluctuations $\Pi_{\rm q}$ predictably
leads to an expansion of the system, see Fig.~\ref{phas}(g).
Consequently, the mean densities of particle number and internal
energy naturally decrease, as shown in Fig.~\ref{phas}(d) and (f).
Additionally, the enhanced fluctuations inhibit the increase of
the central particle density, which is proportional to $\chi^2_0$,
see Fig.~\ref{phas}(c). Moreover, the transition to the predominance
of the rarefied phase is accompanied by an increase in chemical
potential $\nu$, see Fig.~\ref{phas}(a). However, the decreasing
near-linear part of the chemical potential $u$ in Fig.~\ref{phas}(b)
indicates the unstable state.

Physically, we observe that a reduction in the amount of phase-1,
see Fig.~\ref{pnu}, mainly concentrated in the core, results in a
weakened gravitation, which, in turn, facilitates the growth of
quantum fluctuations. On the contrary, the ``massive'' core ensures
the stability of the entire system.

Note that there is confirmation of similar changes in properties
and phase transition under the influence of ``compression'' $\Pi_{\nu}$,
used in \cite{GKN20}:
\begin{equation}
\Pi_{\nu}=-\frac{3}{\xi^3_B}\int_0^{\xi_B}\chi^2(\xi) h(\xi)\xi^2\rmd\xi,
\end{equation}
where $h(\xi)$ is given by (\ref{mu2}).
However, an increase in $\Pi_\nu$ corresponds to a decrease in $\Pi_{\rm q}$
and vice versa. We do not show these results here.

\begin{figure}[h]
\centering
\includegraphics[width=7cm,angle=0]{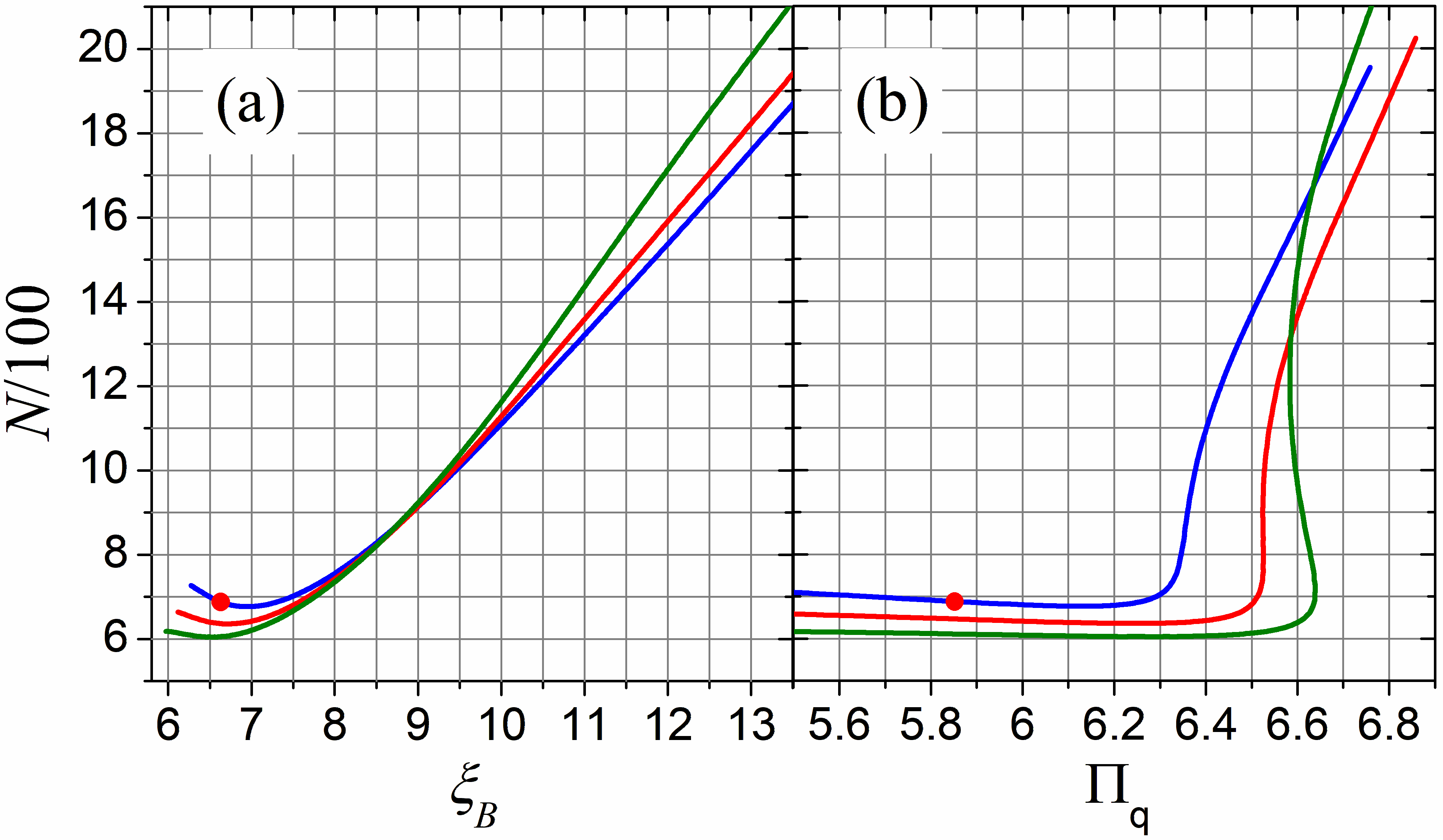} 
\caption{\label{N}Total number of DM particles $N$ as
a function of radius $\xi_B$ (a) and quantum fluctuation
pressure $\Pi_{\rm q}$ (b) for $A=4.5\times10^{-3}$
(blue curves), $A=5.1\times10^{-3}$ (red curves),
$A=5.6\times10^{-3}$ (green curves). The red dot
reflects the configuration of NGC 2366.}
\end{figure}

Valuable information can be also extracted from the mass-radius relation
$M(R)$ by associating the total mass $M=4\pi r_0^3\rho_0 N$ and the radius
$R=r_0\xi_B$ with dimensionless quantities in Fig.~\ref{N}(a). Although
the number $N$ in the case of the unstable rarefied phase dominance tends to 
increase rapidly with radius $\xi_B>8$, we note the presence of a minimum
$N_{\min}$ for relatively dense and compact objects at $\xi_B<8$.
The existence of a minimum point separating the stable and unstable branches
in our approach reproduces predictions from~\cite{BZ19,Ch2018,KT94,SH2018},
in particular for axitons.

This minimum is also seen in Fig.~\ref{N}(b) at $\Pi_{\rm q}\simeq6.15$.
Moreover, the total number (or mass) of particles can exhibit either continuous
or discontinuous behavior during the first-order phase transition, depending
on the value of $A$ in Fig.~\ref{N}(b).

In Fig.~\ref{N}(a) there are also intersections of curves with different $A$.
It is noteworthy that the object configurations with different spatial
distributions that share the same $N$ and $\xi_B$ correspond to the
metastability regions in Fig.~\ref{N}(b).

The BEC DM state of NGC 2366, marked by a red dot in Figs.~\ref{phas} and
\ref{N} at $\Pi_{\rm q}\simeq5.852$, is characterized by the presence of
$x\simeq19.2\%$ of the dense DM in Fig.~\ref{phas}(h), which is concentrated
in $4.7\%$ of the volume. According to our model, it belongs to the (blue)
continuous trajectory, which is classified as {\it supercritical fluid}
with respect to the variations in $(A,\Pi_{\rm q})$.
However, this conclusion relies on spatially averaged values and overlooks
the inhomogeneous DM distribution, as evidenced in Fig.~\ref{sols}
and \ref{press}.  In addition, the phase trajectory of
a macroscopic system can be significantly altered by random or external
influences. In fact, BEC DM of NGC 2366 exhibits a structured, fruitlike
composition, consisting of a dense core, sparse pulp, and crust. These
zones are defined by distinct radial intervals: 1)~\mbox{$r\leq2.83$~kpc},
including a central region for $r\leq0.94$~kpc with a density ratio
$\chi_0^2/\sigma\simeq6.04$; 2)~$2.83\leq r\leq5.55$~kpc;
3)~$5.55\leq r\leq7.79$~kpc, representing the outer layer in the model.

\begin{figure}[h]
\centering
\includegraphics[width=3.8cm,angle=0]{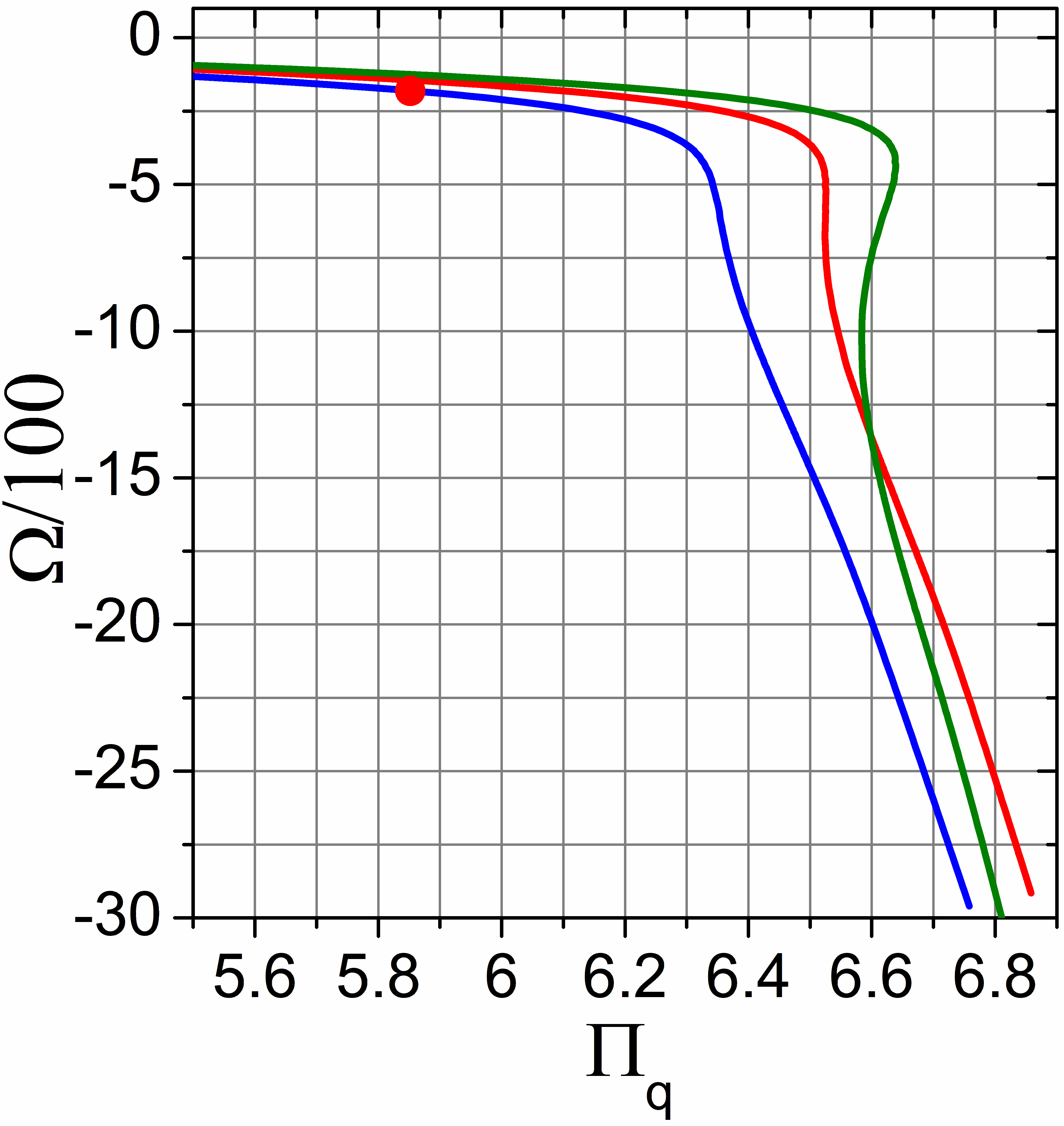} 
\caption{\label{Om}Grand canonical potential $\Omega$ as
a function of quantum fluctuation pressure $\Pi_{\rm q}$
for $A=4.5\times10^{-3}$ (blue curve), $A=5.1\times10^{-3}$
(red curve), $A=5.6\times10^{-3}$ (green curve). The red dot
corresponds to the parameters of NGC 2366.}
\end{figure}

In Fig.~\ref{Om}, the behavior of grand canonical potential,
$\Omega=-(\xi_B^3/3)P$, shows that the DM of NGC 2366 is in
a nearly stable state -- its value of $\Omega$ remains on
a plateau even with small fluctuations in its characteristics.
This may explain the location of the galaxy to the left of 
the critical minimum $N_{\min}$ in Fig.~\ref{N}.

\begin{figure}[h]
\centering
\includegraphics[width=7.7cm,angle=0]{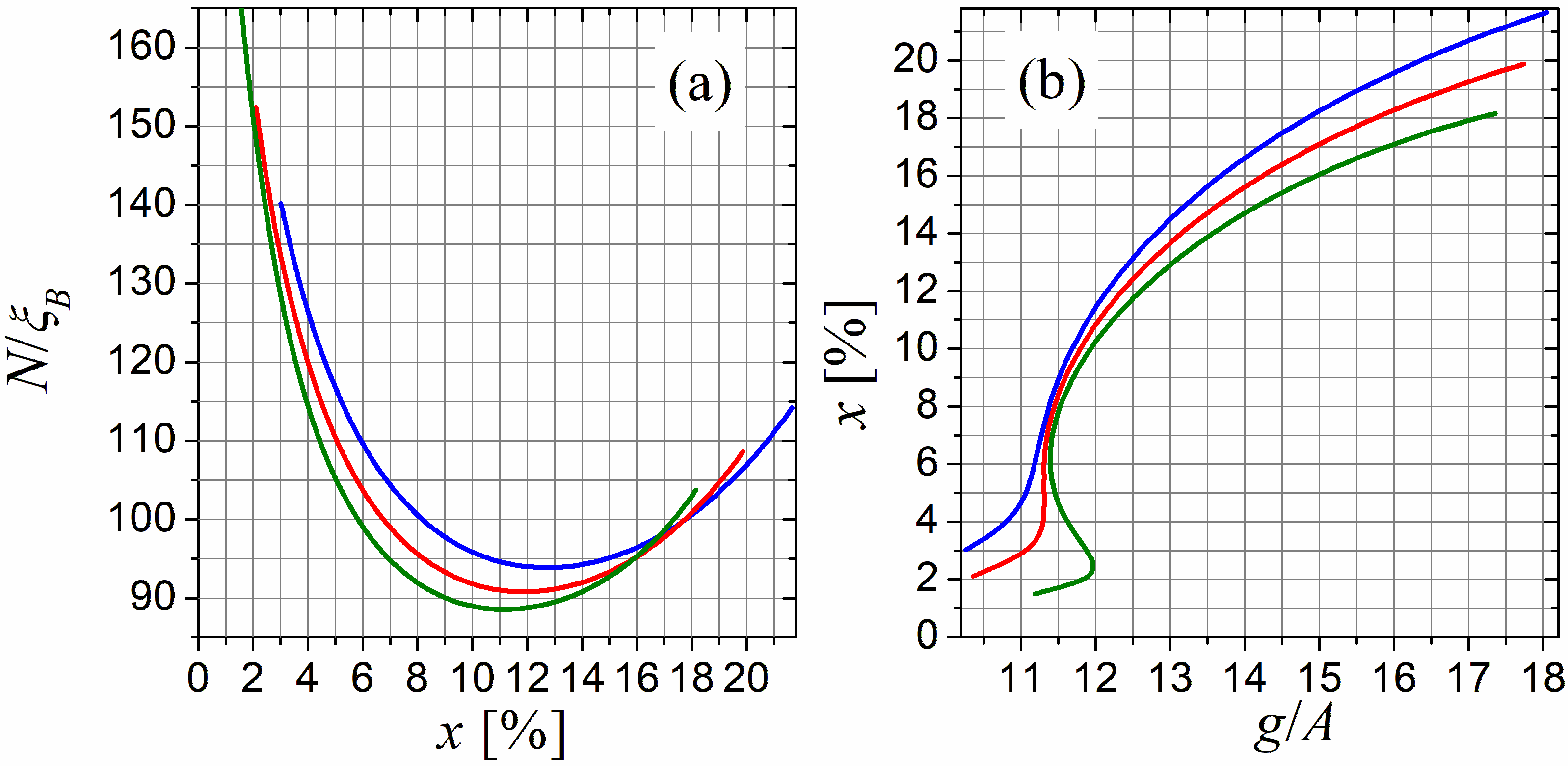} 
\caption{\label{sg}Relations between gravitation properties
and dense DM fraction $x$ for $A=4.5\times10^{-3}$ (blue curves),
$A=5.1\times10^{-3}$ (red curves), $A=5.6\times10^{-3}$
(green curves). Panel (a) illustrates the correlation
between $x$ and the gravitational potential $N/\xi_{B}$.
Panel (b) depicts the dependence of $x$ on dimensionless
surface gravity ${\rm g}/A=N/\xi^2_{B}$.}
\end{figure}

On the other hand, a sharp decrease in $\Omega$ signals a fall
into an unstable configuration.

The parameters that indicate the end of the stability plateau and
the beginning of the instability regime can be associated with
the minimum gravitational potential at the boundary sphere of
radius $\xi_B$ in Fig.~\ref{sg}(a). 
The turning points between stable and unstable configurations,
based on the analysis of three curves, are as follows:
for the blue curve, $x\simeq12.5\%$ and $\Pi_{\rm q}\simeq6.32$;
for the red curve, $x\simeq12\%$ and $\Pi_{\rm q}\simeq6.49$;
and for the green curve, $x\simeq11\%$ and $\Pi_{\rm q}\simeq6.6$.
In Fig.~\ref{sg}(a) the branches of the curves to the right of
the minima show enhanced gravitational attraction due to
a growing fraction of the dense DM component, while the branches
to the left indicate intensified gravitation arising from
the accumulation of DM in the unstable state as the segment $AB$
in Fig.~\ref{pnu}.

In Fig.~\ref{sg}(b), it is shown that the impact of surface gravity,
expressed as ${\rm g}=AN/\xi^2_{B}$, also confirms the presence of a first-order
phase transition in the model, although ${\rm g}$ is not a thermodynamic function
(potential) used to extract critical parameters. We see that its value
becomes larger as $x$ increases.

\section{Conclusion}\label{S4}

We developed a statistical approach to the stationary nonrelativistic BEC DM model
with instanton self-interaction, $V_{\text{inst}}=p_0-2\cos{\chi}$ in terms of
the dimensionless real amplitude $\chi$ of the macroscopic wave function at
$p_0=\pm2$. Introducing the chemical potential $\mu$ and the internal pressure $p$
at zero temperature [see (\ref{muu}) and (\ref{pre})], we classified stable, unstable,
and metastable states of matter using the $p-\mu$ phase diagram in the range
$\chi\in[0; 4.5\pi]$ in Fig.~\ref{pnu}. The phase boundaries $\chi=a_n$ arise from
the zeros of the spherical Bessel function, $j_1(a_n)=0$ with $a_1=0$, defining
intervals of stability $\chi\in(a_{2n}; a_{2n+1})$ and instability
$\chi\in(a_{2n-1}; a_{2n})$ for $n\in\mathbb{N}$. Both the thermodynamic functions
and the excitation spectrum confirm this partitioning, indicating that
the squared sound speed $c^2_s(\chi)$ in (\ref{ssp}) vanishes at $\chi=a_n$, is positive
within stable bands, and negative within unstable ones. In particular,
the small-amplitude regime at $\chi\in[0; \chi_c]$, where
$\chi_c=a_2\simeq4.493409$, is also shown to be unstable, which is consistent
with earlier studies~\cite{Nambo24,KT94}.

We argued that a first-order phase transition is possible between two existing
stable phases with higher $\chi$ in the bands we have defined.
Such amplitudes are usually ignored in the axion theory.

These findings explain why, taking into account self-gravity, the numerical
solutions of the Gross--Pitaevskii--Poisson equations in the Thomas--Fermi
approximation must be piecewise: radial integration necessarily stops at each
``critical'' amplitude $\chi=a_n$, see (\ref{TFeq}).

We apply the BEC model (\ref{feq1})-(\ref{feq2}), which takes into account
self-gravity and quantum fluctuations, to nonperturbatively describe DM in
dwarf galaxies over the range $\chi\in[0; 2.5\pi]$, which spans exactly one
stable and one unstable phase. After prescribing mathematically consistent
initial conditions for the nonlinear spatial evolution (\ref{icon}), we fit
a representative $s$-wave numerical solution to the rotation curve of the
DM-dominated dwarf NGC~2366~\cite{Oh08,Bl08}. The fit yields a dark boson mass
$m \simeq 0.1171 \times 10^{-22}~\text{eV}c^{-2}$, which is comparable to the
mass in \cite{KMT}.  While NGC~2366 exhibits active star formation and
ISM reionization~\cite{Hun01}, we use its DM configuration solely as a
prototypical case of our statistical ensemble of DM distributions in similar
galactic settings.

By requiring the stable phase to form the dense solitonic core of the DM halo,
we compute mean thermodynamic functions for NGC~2366 and a family of hypothetical
analogs under similar boundary conditions (\ref{icon}) and (\ref{bc2}), but
with different core-to-halo mass ratios.
By varying the partial pressure induced by quantum fluctuations $\Pi_{\rm q}$,
we trace the resulting macroscopic characteristics in Fig.~\ref{phas}, which
undergo a first-order phase transition for certain strengths of the gravitational
parameter $A$, the only coupling constant in the model. The revealed jumplike
behavior of the functions is attributed to the ambiguity in the quantum
pressure $\Pi_{\rm q}$. To detect the phase transition, one can also use
the compression~$\Pi_{\nu}$ from~\cite{GKN20} with similar properties
and even the surface gravity value, see Fig.~\ref{sg}(b).

In the case of NGC 2366, however, the DM lies in
the supercritical regime, when its properties vary smoothly. Due to
Fig.~\ref{phas}(h) and \ref{sg}(a), we find that a core exceeding 
the threshold $\gtrsim12\%$ of the total mass enhances gravity and suppresses
quantum fluctuations in the surrounding rarefied halo, while the internal
pressure $p$ ensures core stability. Concretely, NGC~2366’s DM dense core
contains $\approx19\%$ of the DM mass but occupies only $\approx4.7\%$ of
its volume.

We define the threshold of the core mass from the gravitational potential extremum
in Fig.~\ref{sg}(a), comparing it with the end of the stability plateau of the grand
canonical potential $\Omega$ in Fig.~\ref{Om}. The extremum existence means that
the gravitational potential values are repeated for different configurations, thereby
causing ambiguity in thermodynamic potential density $\omega$ in Fig.~\ref{omx}.

Moreover, in the mass-radius diagram in Fig.~\ref{N}(a), which shows stable and
unstable branches analogous to those in~\cite{BZ19,Ch2018,KT94,SH2018},
the DM configuration for NGC~2366 lies firmly on the stable branch, consistent
with the observation. On the other hand, the internal mechanisms of the model
are effective in addressing the problem of gravitational collapse~\cite{Ch2016,Khl}
of the BEC DM halo, see e.g. \cite{Guzman,H2019,H2014}.

Thus, the thermodynamic signature provides a novel diagnostic for distinguishing 
stable and unstable DM configurations in galactic halos. Viewed from an ergodic
perspective, it offers an alternative to time-dependent approaches, and a direct
comparison between the two deepens the understanding of the DM halo evolution.

Of course, the physical picture goes beyond macroscopic properties.
In this regard, we mention at least the ability of DM particles in
phase-1, i.e. at $\chi_c<\chi\lesssim2\pi$, to form composites --
two-particle dimers -- during resonant scattering, which follows from \cite{GN23}.

\acknowledgements

This work is supported by the Ukrainian-Swiss project No. IZURZ2-224868
``Compact star-forming galaxies and their impact on cosmic reionization
and cosmology''.


\end{document}